\documentclass[%
 aip,
 amsmath,amssymb,
reprint,%
]{revtex4-1}

\usepackage{graphicx}
\usepackage{dcolumn}
\usepackage{bm}
\usepackage[utf8]{inputenc}
\usepackage[T1]{fontenc}
\usepackage{mathptmx}
\usepackage{etoolbox}

\newcommand{\comment}[1]{}

\newcommand{\vecunder}[1]{\pmb{#1}}
\newcommand{\matunder}[1]{\pmb{#1}}

\begin{document}


\title{Methods for Inferring Interaction Potentials from Cross-Linking Mass Spectrometry Data}

\author{B\"{o}rries von Seggern}
\affiliation{%
    Department of Physics, Freie Universit\"{a}t Berlin, 14195 Berlin, Germany
}

\author{Mohsen Sadeghi}
\email[Corresponding author: ]{mohsen.sadeghi@fu-berlin.de}
\affiliation{%
    Department of Mathematics \& Computer Science, Freie Universit\"{a}t Berlin, 14195 Berlin, Germany
}
\affiliation{%
    Zuse Institute Berlin, 14195 Berlin, Germany
}

\date{\today}

\begin{abstract}
Cross-linking mass spectrometry (XL-MS) has emerged as a powerful quantitative technique for probing intra-protein structural information as well as protein-protein interactions at an unprecedented scale. XL-MS data yield information on the pairwise spatial proximity of proteins through inter-molecular linkers. However, systematic methods for translating these experimental measurements into computational models on the same scale remain limited. Specifically, determining interaction potentials from experimental observables is crucial in particle-based modeling. Existing methods are predominantly focused on directly fitting radial distribution functions (RDFs), while numerous observables, e.g. coordination numbers, which are functionals of the RDF, cannot be uniquely inverted. In addition, complexities such as phase separation and mixing of multiple components make these methods largely unsuitable. In this work, we develop a framework for parameterizing interaction potentials from such observables in potentially phase-separated mixtures, demonstrated by parameterizing models from XL-MS results. We establish a connection between this problem and the inverse Henderson problem and adapt algorithms such as Iterative Boltzmann Inversion and Iterative Monte Carlo to its numerical solution. We derive exact and low-density limit gradient approximations and propose two new algorithms based on an adaptation of the predictor-corrector~framework. In total, we evaluate several optimization algorithms on biologically realistic ten-component test systems. We demonstrate that for homogeneous fluids, all methods achieve exceptional efficiency and accuracy. Critically, we further demonstrate successful parametrization in a challenging three-phase system. Here, three algorithms, namely Adam and gradient descent employing the low-density derivative as well as Newton's method with the exact gradient, reliably recover the correct parameters. These results establish a clear pathway from XL-MS experiments to coarse-grained protein models for systems where phase separation governs biological function, potentially enabling new investigations of biomolecular condensates and protein aggregation.
\end{abstract}

\maketitle

\section{\label{sec:Intro}Introduction}

Non-homogeneous fluid systems are of crucial importance across a range of fields and applications in both soft matter and biology, with a prominent example being the mixture of proteins within the cytoplasm of living cells. Constructing particle-based computational models of these systems necessitates reliable means of determining the matrix of inter-particle interactions. The predictive power of such models, at any scale, is directly tied to how realistic these interaction potentials are. Despite this, methods for parametrizing coarse-grained (CG) interaction potentials in fluids have predominantly focused on homogeneous systems and fine-grained structural data, such as those obtained from scattering techniques.~\cite{LwrOriginalWork, PCIBISchommers} Extensive literature exists on determining interaction potentials to reproduce radial distribution functions (RDFs). Numerous methods have emerged for solving this problem in structural coarse-graining~\cite{NewIBIReithDirkPutz, OriginalIMC, IHNCDelbary}, with applications in complex scenarios, such as proteins on the surface of biomembranes.~\cite{AggregationMohsen, EntropyMohsen} However, the extension of these methods to work with less detailed structural information remains challenging, as classical approaches typically rely on direct knowledge of the target RDFs.

One source of such coarse-grained observables in structural biology is cross-linking mass spectrometry (XL-MS). In XL-MS, samples are exposed to a bifunctional chemical reagent, aptly named the linker, that can form selective covalent bonds between pairs of residues, predominantly lysines, creating cross-linked pairs that can be identified through mass spectrometry. Information about spatially-proximate residues having been thus linked not only uncovers 3D strutural information about single proteins but also reveals if such residue pairs existed between neighboring proteins. Due to advances in cross-linker chemistry and highly automated spectrometric analysis, XL-MS surveys have reached unprecedented scale, enabling even proteome-wide investigations (for a comprehensive review of XL-MS techniques see e.g. Ref.~\onlinecite{XLMS_Review}). This capability to probe protein association at the large scale makes XL-MS valuable for gaining insight into proteome-wide phenomena, such as \emph{in vivo} protein localization in the crowded cytosolic environment.~\cite{MitochaondriaLocalFanLiu}

Protein phase separation, particularly in biomolecular condensates, has gained considerable attention recently, since the discovery of their involvement in a range of biological processes.~\cite{FirstSignallingCondensates, NucleolarSubcompartmentsCondensates} Furthermore, phase separation is also implicated in disease pathology such as Alzheimer's disease, where amyloid plaques form through protein aggregation. Despite this biological significance and the growing availability of quantitative XL-MS data, systematic methods for interpreting XL-MS results, for instance, by parameterizing CG protein models remain limited.~\cite{IntegrativeMD_XLMS}

In the context of multicomponent protein mixtures, XL-MS data is to be interpreted as an ensemble average of protein-protein contacts. In this work, we therefore develop and evaluate systematic methods for parameterizing interaction potentials from such CG~observables in potentially phase-separated mixtures. We establish a general framework connecting experimental observables to interaction parameters and propose several optimization algorithms adapted from classical methods. We demonstrate the performance of these methods on both homogeneous and multiphase model systems inspired by biology, with target observables derived from a plausible interpretation of XL-MS results.
 
The remainder of this paper is organized as follows. Section~\ref{sec:Theory} presents the theoretical framework, optimization algorithms, and the CG~model for XL-MS results. Section~\ref{sec:Exp} details the computational methods and simulation protocols. Section~\ref{sec:Results} presents results for both homogeneous and inhomogeneous systems, comparing the performance of different optimization strategies. Finally, Section~\ref{sec:Conclusions} summarizes our findings and discusses implications for CG~modeling of complex biological systems.

\section{\label{sec:Theory} Theory}
\subsection{\label{sec:Theory:OptFramework}Coarse Grained Optimization Framework}
Structural observables obtained from experiments provide a natural foundation for developing CG~models in soft matter and biomolecular systems \cite{MembraneMohsen}. In the following, a general theoretical framework is outlined that connects these observables to the underlying interaction parameters and establishes a basis for their optimization in phase-separated multicomponent systems.

\paragraph{Model and Observables}
We consider a mixture of~$M$ species in the canonical ensemble at temperature~$T$, volume~$V$, and particle numbers~$\vecunder{N}=(N_a)_{a=1}^M$.  
The full Hamiltonian of the system is given by
\begin{equation}
    \mathcal{H}(\vecunder{r}^N, \vecunder{p}^N; \vecunder{\varepsilon}) 
    = \sum_{i=1}^N \frac{\vecunder{p}_i^2}{2m_i}
      + \sum_{i=1}^N\sum_{j> i}^N 
      u_{a_i b_j}\!\left(r_{ij}; \vecunder{\varepsilon}\right),
    \label{eqn:Hamiltonian}
\end{equation}
where~$r_{ij}$ is the distance between particles~$i$ and~$j$. Additionally, $N$~designates the total number of particles, whereas~$a_i$ and~$b_j$ specify the species type of particles~$i$ and~$j$, respectively. The pairwise additive potentials $u_{ab}(r;\vecunder{\varepsilon})$ depend on a dimensionless parameter vector $\beta\vecunder{\varepsilon} \in \mathbb{R}^P$, where $\beta = (k_{\text{B}}T)^{-1}$ denotes the inverse temperature and $k_{\text{B}}$ is the Boltzmann constant.

We focus on a set of $Q=\frac{1}{2}M(M+1)$~measurable observables~$\vecunder{z}=(z_i)_{i=1}^Q$, where indices~$i$ correspond to pair types unique up to ordering ($ab=ba$). For now, we assume that each~$z_i$ can be expressed as a functional of the RDF~$g_i(r)$,
\begin{equation}
    z_i = \int f_i(r)\, g_i(r)\,dr \; ,
    \label{eqn:GeneralObservable}
\end{equation}
where $f_i(r)$~denotes a weighting function that defines the mapping between the RDF and the experimental observable (e.g., a coordination number).  
The general dependence of the observables on the parameters is then written as
$
    {\vecunder{z}(\vecunder{\varepsilon})~=~\vecunder{z}\!\left[\vecunder{u}(\vecunder{\varepsilon})\right]} \;
$.
Here, we cast the parametrization problem as determining a parameter set~$\vecunder{\varepsilon}$ that minimizes a loss function~$\mathcal{L}\left(\vecunder{z}(\vecunder{\varepsilon}),\vecunder{z}^\ast\right)$ measuring the deviation between model observables and target (experimental) data~$\vecunder{z}^\ast$. In the following, we define this optimization problem as the Inverse Pair Problem due to its conceptual analogy to the Inverse~Henderson~Problem~(IHP).

\paragraph{Relation to the Inverse Henderson Problem}
Scattering techniques provide access to RDFs on atomistic length scales, and a large body of work is dedicated to determining interaction potentials that reproduce RDFs. 
This problem, termed the IHP, can be understood by defining a (forward) Henderson map~$\mathcal{G}: u \mapsto g$ as the relation between interaction potential~$u$ and corresponding RDF~$g$.~\cite{IHNCDelbary} For systems governed exclusively by pairwise additive interactions, the Henderson Theorem states that~$u$ is uniquely determined by~$g$ (up to an additive constant) at fixed~$T$ and~$\rho$.~\cite{HendersonTheorem} As outlined by Delbary~\textit{et al.},~\cite{IHNCDelbary} determining~$u$ from~$g$ can thus be seen as seeking an inverse map~${\mathcal{G}^{-1} : g \mapsto u}$. 

For mixtures this can be generalized to the vectorial map  
\begin{equation}
    \vecunder{\mathcal{G}}:\vecunder{u}\mapsto\vecunder{g}
\end{equation}
relating a set of potentials to the corresponding partial RDFs. The given parametrization problem can then be written as
\begin{equation}
    \vecunder{z}^\ast
    = \vecunder{z}\left[
        \vecunder{\mathcal{G}}\left(\vecunder{u}(\vecunder{\varepsilon})\right)
      \right]\ .
\end{equation}
However, the functional dependence~$z_i[\vecunder{g}]$ is not unique, such that no single $g$ can be identified as a target for IHP methods. Consequently, IHP methods that approximate~$\vecunder{\mathcal{G}}^{-1}$ cannot be directly used to invert the composite~map~$\vecunder{z}\circ\vecunder{\mathcal{G}}$, necessitating new solution strategies.

\paragraph{Forward Evaluation}
Two methods for evaluating the forward map~$\vecunder{\mathcal{G}}(\vecunder{u})$ are most relevant for this work. The first method is molecular dynamics (MD), in which the configuration space is sampled directly. In the given context, the main advantage of MD simulations is the ability to describe multiphase systems, while the main disadvantage is MD's high computational cost compared to the second approach, i.e. using Integral Equation Theories~(IET).

IETs such as the hypernetted~chain~(HNC)~approximation offer an alternative evaluation method for homogeneous systems. IETs start with the Ornstein-Zernike~relation
\begin{equation}
h_{ab}\left(|\vecunder{r}|\right) = c_{ab}\left(|\vecunder{r}|\right) + \sum_{\lambda=1}^M \rho_\lambda \int h_{a\lambda}\left(|\vecunder{r}'|\right) c_{\lambda b}\left(|\vecunder{r}-\vecunder{r}'|\right) \,d\vecunder{r}' \ ,
    \label{eqn:IETOZrelation}\
\end{equation}
relating~$h_{ab}(r)=g_{ab}(r)-1$ to the set of so-called direct correlation functions~$c_{ab}(r)$. The Ornstein-Zernike relation can indeed be seen as the definition of~$c_{ab}(r)$, and when combined with a closure relation, such as the HNC~approximation,
\begin{equation}
\ln{(g_{ab}(r))} = -\beta u_{ab}(r) - c_{ab}(r) + h_{ab}(r) \ ,
    \label{eqn:HNC_Closure}
\end{equation}
it forms a system of equations that can be solved numerically to approximate~$h_{ab}(r)$. The HNC~approximation is exact to order~$\rho^2$ and remains particularly accurate for potentials with soft repulsive cores, such as the potentials described below.

Solutions of the HNC~approximation can be numerically obtained at a fraction of the computational cost associated with MD. However, the description of phase-separated systems using these methods is nontrivial since solutions have to be obtained for each individual phase in the system. Thus, the phase structure of the system has to be identified in advance making these methods impractical for describing complex multiphase systems.

\subsection{\label{sec:Theory:OptAlgos} Optimization Algorithms}
Irrespective of the method used for obtaining estimates of the~$\vecunder{z}$ functionals, minimizing the loss function~$\mathcal{L}\left(\vecunder{z}(\vecunder{\varepsilon}),\vecunder{z}^\ast\right)$ constitutes a complicated non-convex optimization problem. The numerical algorithms required for this process are listed in the following. We remind the reader that the optimization is always done with respect to the vector of potential parameters~$\vecunder{\varepsilon}$.

\paragraph{Gradient Descent}
Gradient Descent~(GD) methods provide a direct approach to minimizing~$\mathcal{L}(\vecunder{z}(\vecunder{\varepsilon}))$. In plain GD, a minimum of~$\mathcal{L}\left(\vecunder{z}\left(\vecunder{\varepsilon}\right)\right)$ can be attained by updating the parameter set~$\vecunder{\varepsilon}_k$ according to
\begin{equation}
    \beta\vecunder{\varepsilon}_{k+1} = \beta\vecunder{\varepsilon}_k - \alpha \nabla_{\beta\varepsilon}\mathcal{L}\left(\vecunder{z}\left(\vecunder{\varepsilon}_k\right)\right)
\end{equation}
at each step~$k$. Typically, the step size parameter~$\alpha$ is chosen to give appropriately small updates, while~$\nabla_{\beta\varepsilon}\mathcal{L}\left(\vecunder{z}\left(\vecunder{\varepsilon}_k\right)\right)$ denotes the gradient of~$\mathcal{L}\left(\vecunder{z}\left(\vecunder{\varepsilon}_k\right)\right)$ with respect to~$\beta \vecunder{\varepsilon}$, evaluated at~$\vecunder{\varepsilon}_k$. In general, evaluating~$\nabla_{\beta \vecunder{\varepsilon}}\mathcal{L}$ requires the Jacobian
\begin{equation}
    \matunder{J}_z = \left(\frac{\partial z_i}{\partial \varepsilon_j}\right)_{i,j} \; .
\end{equation}
In machine learning, GD is typically performed using noisy estimates of the true gradient. In these stochastic GD~applications, adaptive optimizers such as Adam can improve convergence by adjusting step sizes based on the gradient history.~\cite{AdamPaper} 

\paragraph{Derivative Approximations}
We propose two approaches for estimating the Jacobian~$\matunder{J}_z$. The first method is inspired by the Iterative~Monte~Carlo method, which employs an analogous Jacobian in solving the IHP.~\cite{OriginalIMC} In the canonical ensemble, the exact derivative of an observable~$O$ with respect to a parameter~$\varepsilon$ of the Hamiltonian~$\mathcal{H}(\varepsilon)$ is given by
\begin{equation}
    \partial_\varepsilon \langle O \rangle = -\beta\left( \left\langle O\, \left(\partial_\varepsilon \mathcal{H} \right)\right\rangle - \langle O \rangle \left\langle \partial_\varepsilon \mathcal{H} \right\rangle \right)\ .
    \label{eqn:ExactDerivative}
\end{equation}
Here, $\langle\cdot\rangle$~denotes the canonical ensemble average of an observable~$O(\vecunder{r}^N, \vecunder{p}^N)$,
\begin{equation}
    \langle O\rangle = \frac{\int O\left(\vecunder{r}^N,\,\vecunder{p}^N\right) e^{-\beta \mathcal{H}\left(\vecunder{r}^N,\,\vecunder{p}^N\right)}\,\text{d}\vecunder{r}^N\,\text{d}\vecunder{p}^N}{\int e^{-\beta \mathcal{H}\left(\vecunder{r}^N,\,\vecunder{p}^N\right)}\,\text{d}\vecunder{r}^N\,\text{d}\vecunder{p}^N}\ .
\end{equation}
Using Eq.~\ref{eqn:ExactDerivative} implicitly relies on higher-order correlation functions (such as four-body densities), and thus, this expression is not accessible from IET results alone and needs to be sampled through MD~simulations. While the derivative is exact in principle, its evaluation is limited by finite sampling in practice, motivating the use of stochastic GD~algorithms such as Adam.

An analytically tractable alternative exists in the low-density limit (LDL). In the limit~$\rho\to0$, the RDF approaches~$g(r)=e^{-\beta u_i(r)}$. As a slight generalization of this result, it can be assumed that the cavity distribution function~$y_i(r)=e^{\beta u_i(r)} g_i(r)$ becomes insensitive to changes in~$u_i(r)$. This yields the LDL model
\begin{equation}
    g_i(r) = e^{-\beta u_i(r)} \overline{y}_i(r)
    \label{eqn:LDL-RDF}
\end{equation}
with constant~$\overline{y}_i(r)$. Differentiating this expression gives
\begin{equation}
    \frac{\partial g_i(r)}{\partial \varepsilon_j} = -\beta\, g_i(r)\,\frac{\partial u_i(r)}{\partial \varepsilon_j}\; ,
    \label{eqn:LDL_Derivative}
\end{equation}
which leads to the  approximation
\begin{equation}
    \frac{\partial z_i}{\partial \varepsilon_j} = \int f_i(r)\frac{\partial g_i(r)}{\partial \varepsilon_j}\,dr  =  -\beta\int f_i(r)g_i(r)\,\frac{\partial u_i(r)}{\partial \varepsilon_j}\,dr \ .
\end{equation}
This derivative in the low-density limit can be evaluated directly from knowledge of~$g_i(r)$ and is thus available with both MD~and HNC~results. It should be noted that more accurate analytical gradient approximations, e.g. using the HNC~approximation can be derived.~\cite{IHNCDelbary} However, since these are based on IETs, we found these approximations to be computationally intractable in inhomogeneous systems.

\paragraph{Predictor-Corrector Methods}
A powerful approach for deriving IHP methods is offered by the predictor-corrector~framework.~\cite{PCIBISchommers,  LwrOriginalWork, RaoRajagopalanRHNCinversion3D, LawBuzzaRHNCinversion2D} In predictor-corrector~methods, an approximate model of the forward Henderson map~$\vecunder{\mathcal{G}}_k$, is analytically inverted in the predictor step to obtain an updated potential~$\vecunder{u}_{k+1}$ by imposing
$
    \vecunder{\mathcal{G}}_k(\vecunder{u}_{k+1}) \stackrel{!}{=} \vecunder{g}^\ast\,.
$
In a subsequent corrector step, the predictor model~$\vecunder{\mathcal{G}}_{k+1}$ is then updated using the updated potential~$\vecunder{u}_{k+1}$. A particularly illustrative example of such a method is Iterative Boltzmann Inversion~(IBI). While the more general derivation of IBI is based on the potential of mean force,~\cite{NewIBIReithDirkPutz} in a single-component system governed exclusively by pairwise-additive two-body forces, IBI can also be derived as a predictor-corrector~method where the predictor model is the LDL approximation~$\mathcal{G}_{k}(u)=e^{-\beta u}\overline{y}_k$.~\cite{PCIBISchommers} Inverting~$\mathcal{G}_k(u_{k+1})=g^\ast$ here yields the IBI~update
\begin{eqnarray*}
    u_{k+1}(r) 
    &= -\beta^{-1}\ln\!\left(\frac{g^\ast(r)}{\overline{y}_k(r)}\right)
     = u_k(r) - \beta^{-1}\ln\!\left(\frac{g^\ast(r)}{g_k(r)}\right)\ .
\end{eqnarray*}
The corrector step then consists of generating a new estimate of~$g_k$ (and thus~$\overline{y}_k$) through a simulation with the updated potential~$u_{k+1}$.

However, the present problem differs fundamentally from the IHP. Since the projection~$z_i[g_i]$ cannot be inverted, possible predictor models~$\vecunder{z}_k(\vecunder{\varepsilon})=\vecunder{z}\circ\vecunder{\mathcal{G}}_k$ obtained from approximate~$\vecunder{\mathcal{G}}$ can not be inverted explicitly either. Consequently, the predictor step has to be adapted to this new problem. Here, we propose an adaptation inspired by trust-radius algorithms. The proposed predictor step is given by finding a local minimum of~$\mathcal{L}\left(\vecunder{z}_k \left(\vecunder{\varepsilon}\right) \right)$ employing an approximate predictor model~$\vecunder{z}_k \left(\vecunder{\varepsilon}\right)$ within a ball of radius~$r_\varepsilon$ around the {current iterate}
\begin{equation*}
    \vecunder{\varepsilon}_{k+1} \in \{\beta\vecunder{\epsilon}\in K_{r_\varepsilon}\left(\vecunder{\varepsilon}_k\right) : \vecunder{\epsilon} \text{ is a local minimum of } \mathcal{L}\left(\vecunder{z}_k\left(\vecunder{\epsilon}\right)\right)\}\;,
\end{equation*}
\begin{equation*}
    K_{r_\varepsilon}\left(\vecunder{\varepsilon}_k\right) = \left\{\beta\vecunder{\epsilon} \in \mathbb{R}^{P} :  r_\varepsilon^2 \geq \frac{1}{P}||\beta\vecunder{\epsilon} - \beta\vecunder{\varepsilon}_k||^2_2 \right\}\,.
\end{equation*}
In a subsequent corrector step the predictor map~$\mathbf{z}_k(\vecunder{\varepsilon})$ is then updated using results from a more accurate evaluation at~$\vecunder{\varepsilon}_{k+1}$ e.g. an MD simulation or HNC~solution. 

We propose two predictor-corrector~methods that parallel successful IHP methods. The first, termed \textit{trust-LDL}, can be seen as a natural extension of IBI.~\cite{PCIBISchommers, NewIBIReithDirkPutz} Here, the predictor model at iteration~$k$ is given by the LDL approximation (equation~\ref{eqn:LDL-RDF}), such that the observable~$z_i$ is estimated by
\begin{equation}
    \left(\vecunder{z}_k\left(\vecunder{\varepsilon}\right)\right)_i = \int_0^\infty f_i(r) e^{-\beta u_i(r;\vecunder{\varepsilon})} {y}_{i,k}(r) \,dr \;,
    \label{eqn:LDL-Predictor}
\end{equation}
with an estimate of the cavity distribution function~${y}_{i,k}(r)$ that is updated in the subsequent corrector step.

The second approach is similar to Newton's method, such as employed in Iterative Monte Carlo.~\cite{OriginalIMC} Letting~$\matunder{J}_z\left(\vecunder{\varepsilon}_k\right)$ denote the Jacobian evaluated at~$\vecunder{\varepsilon}_k$, the predictor model is given by the linearization
\begin{equation}
    \vecunder{z}_k\left(\vecunder{\varepsilon}\right) = \vecunder{z}\left(\vecunder{\varepsilon}_k\right) + \matunder{J}_z\left(\vecunder{\varepsilon}_k\right) \cdot \left(\vecunder{\varepsilon} - \vecunder{\varepsilon}_k\right) \, .
    \label{eqn:Linear-Predictor}
\end{equation}
In the corrector step, the Jacobian can be evaluated using either the LDL approximation or exact derivative, which gives two methods that will be referred to as the \textit{Newton (LDL)} and \textit{Newton (Exact)} methods.

\subsection{\label{sec:Theory:CGModel} CG Model for XL-MS Results}
In structural biology, XL-MS results are often interpreted as constraints on the spatial proximity of protein pairs. Similarly, we assume that the experimentally determined number of cross-links can be used to deduce the equilibrium pair number~$z_{ab}$, which we define as the average number of particles within some cutoff distance~$r^M_{ab}$. In the framework established in Section~\ref{sec:Theory:OptFramework}, this corresponds to choosing the weighting function 
\begin{equation}
    f_{ab}(r) = \frac{N_{a}N_{b}} {V (1+ \delta_{ab})} 4 \pi r^2 \Theta(r^M_{ab} - r) \ ,
\end{equation}
where~$\Theta(r)$ denotes the Heaviside function, $N_a$~and~$N_b$ are the particle numbers, $V$~is the system volume, and the Kronecker-Delta~$\delta_{ab}$ corrects for double-counting between like species. The total pair number is then given by
\begin{equation}
     z_{ab} = \frac{N_{a}N_{b}} {V (1+ \delta_{ab})} \int_0^{r^M_{ab}} 4 \pi r^2 g_{ab}(r) \,dr \ .
    \label{eqn:ConnectionNAB-gR}
\end{equation}

\begin{figure}
    \centering
    \includegraphics[width=\linewidth]{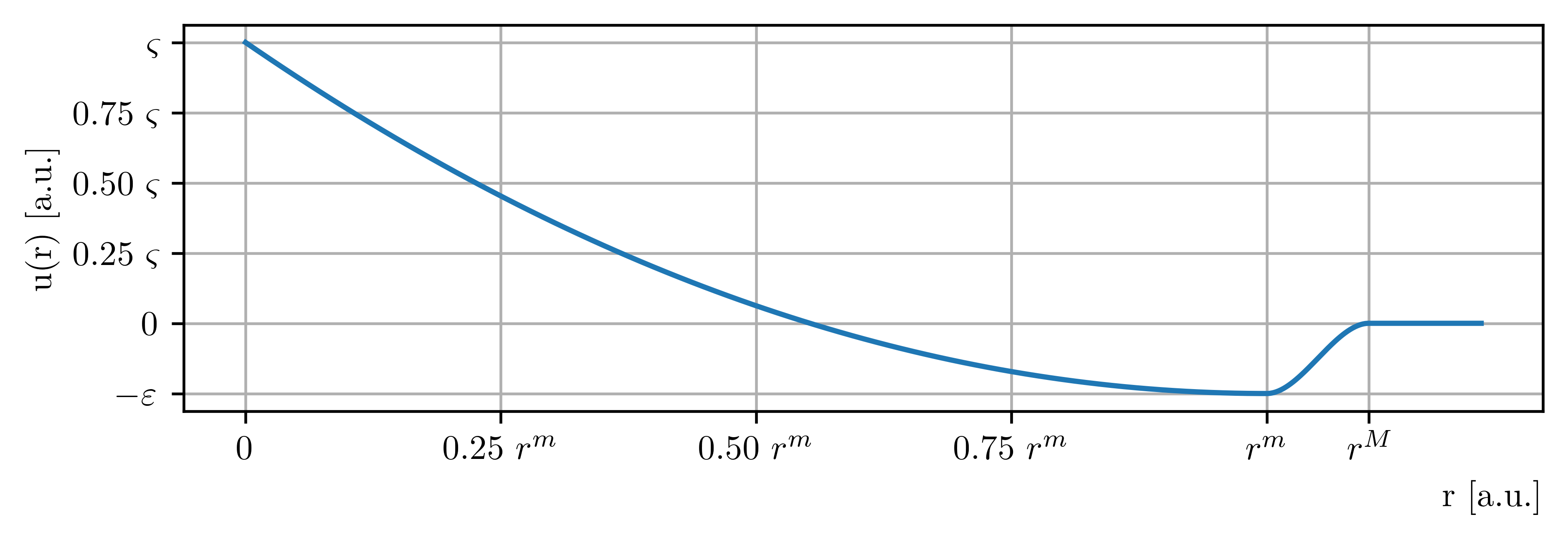}
    \caption{Representative plot of the potential given by eqn.~\ref{eqn:ApproxPMF}}.
    \label{fig:PotPlot}
\end{figure}

To model the protein mixture, we employ a CG description where proteins are treated as spherically symmetric particles in an implicit solvent. The system is assumed to be described by the canonical ensemble and the effective interactions between proteins are modeled using a pairwise-additive soft-core potential. The employed interaction potentials~$\beta u_{ab}(r)$ are similar to those commonly employed in Dissipative Particle Dynamics (DPD),~\cite{HoogerFirstDPD, DPDsolidFoundation} and given by
\begin{equation}
    u_{ab}(r) = \begin{cases}
    \left(\varsigma_{ab} + \varepsilon_{ab}\right) \left(1- \frac{r}{r_{ab}^m}\right)^2 - \varepsilon_{ab} & r \leq r_{ab}^m \\
    - \varepsilon_{ab} \frac{\left(r_{ab}^M - r\right)^2 \left(r_{ab}^M - 3r_{ab}^m + 2r \right)}{\left(r_{ab}^M - r_{ab}^m \right)^3} & r_{ab}^m < r \leq r_{ab}^M \\
    0 & r_{ab}^M < r \\
    \end{cases} \, ,
    \label{eqn:ApproxPMF}
\end{equation}
where~$r_{ab}^m$ defines the range of the repulsive core and~$r_{ab}^M$ denotes the interaction cutoff radius. An representative plot of the potential is displayed in Figure~\ref{fig:PotPlot}. The parameters~$\varsigma_{ab}$ and~$\varepsilon_{ab}$ control the repulsion strength and the depth of the attractive well, respectively. In the context of the outlined optimization problem, the attraction strengths~$\varepsilon_{ab}$ serve as the tunable parameters~$\vecunder{\varepsilon}$ to be fitted, while geometric parameters ($r_{ab}^m,\, r_{ab}^M$) and repulsion strengths ($\varsigma_{ab}$) are treated as fixed constants determined by protein size.

For the given optimization problem, the pair numbers~$z_{ab}$ serve as target observables. Motivated by the structure of the Boltzmann prefactor of each~$z_i$, we employ an offset logarithmic loss function,
\begin{equation}
    \mathcal{L}_{\text{log}}\left(\vecunder{z}\left(\vecunder{\varepsilon}\right)\right) = \frac{1}{Q} \sum_{i=1}^{Q} \left(\ln{\left(\frac{z_i\left(\vecunder{\varepsilon}\right)}{z_i^\ast} + \iota\right)}\right)^2\;,
\label{eqn:MSLogLoss}
\end{equation}
which also treats relative errors uniformly across observables of different magnitudes. The offset~$\iota$ is chosen to be~$\iota=10^{-8}$ and prevents issues when~$z_i\left(\vecunder{\varepsilon}\right)$ is numerically evaluated to be zero (e.g. in an MD simulation), while being negligible compared to any target~$z_i$.

\section{\label{sec:Exp} Computational Details}
\subsection{\label{sec:Exp:Systems}Systems}
The systems investigated in this study are largely based on the tegument of human cytomegalovirus for which accurate protein expression numbers were available.~\cite{HCMmodelMohsen} The reference systems used in Section~\ref{sec:Results:Homogeneous} and~\ref{sec:Results:InHomogeneous} comprise the ten most abundant tegument proteins, simulated at volumes of~$V=3\times10^6~\text{nm}^3$ and~$V=9\times10^6~\text{nm}^3$ for the multiphase and homogeneous systems, respectively. The multiphase system's volume is comparable to the tegument volume,~\cite{3DHCMmodel1999, AtomicCapsidModel} with comprehensive details provided in Appendix~\ref{chp:AppendixReferenceSimulations}.

The potential defined in equation~\ref{eqn:ApproxPMF} requires four parameters for each species pair~$ab$. Beyond the primary objective of determining~$\varepsilon_{ab}$, additive radii were employed such that~${r_{ab}^m=r_{a}^m + r_{b}^m}$, with the larger radius defined by~${r_{ab}^M=1.1\,r_{ab}^m}$. The individual radii were set to the radius of gyration of the corresponding protein structure for the ten-component reference systems. These protein structures were generated by \textsc{AlphaFold2}.~\cite{AlphaFold2} To prevent particles from collapsing onto another, a core repulsive strength of~$\varsigma_{ab}=50\,k_\text{B} T$ was consistently applied across all particles in this study.

\subsection{\label{sec:Exp:SimProtocol}Simulation Protocol}
Simulations were conducted using periodic boundary conditions and the minimum image convention in cubic simulation boxes with an implementation of the potential in \textsc{openMM~8.1.0}.~\cite{openMM8} All proteins were assigned a mass of~${m=30000~\text{u}}$, and the equations of motion were integrated using the \textsc{LangevinMiddleIntegrator} at a temperature of~${T=298~\text{K}}$ employing friction coefficient~${\gamma=850~\text{ps}^{-1}}$ and integration timestep~${\Delta t=10~\text{ps}}$. Unless otherwise stated, initial particle positions were randomly assigned on an equidistant grid, with a subsequent energy-minimization lasting for up~to~$10^6$ steps or until residual forces were within~$10^{-3}~\text{kJ}~\text{mol}^{-1}~\text{nm}^{-1}$. Following minimization, the system was equilibrated for a duration~$t_{\text{eq}}$ and the trajectory sampled every~$100~\text{ns}$ for a time of~$t_{\text{prod}}$. To guide particles into forming a \textit{single} condensed phase, a weak harmonic potential
\begin{equation}
    V_{\text{ext}}(\vecunder{r}) = -\frac{k_\text{ext}}{2} \left(\frac{2|\vecunder{r}|}{L}\right)^2
    \label{eqn:ExtEqForce}
\end{equation}
was applied to all particles during energy minimization and the initial 25~\% of the equilibration period. Here, a constant of~$k_\text{ext}=0.1\,k_\text{B} T$ was employed, while $L$~denotes the length of the simulation box.

\subsection{\label{sec:Exp:SimParams}Simulation and Optimization Parameters}

Reference simulations were performed to obtain target~$\vecunder{z}^\ast$ to fit using the proposed algorithms, which is outlined in Section~\ref{sec:Results:Homogeneous} and~\ref{sec:Results:InHomogeneous}. These simulations were initialized as described above and~$t_\text{eq}=3~\text{ms}$ with~$t_\text{prod}=3~\text{ms}$ was used for the multiphase systems, whereas~$t_\text{eq}=1~\text{ms}$ and~$t_\text{prod}=3~\text{ms}$ were employed for the homogeneous reference simulations.

The MD simulations used for optimization in Section~\ref{sec:Results:InHomogeneous} employed equilibration and production times of~$t_\text{eq}=0.1~\text{ms}$ as well as~$t_\text{prod}=0.5~\text{ms}$, respectively. To reduce the required equilibration time in each iteration, the particle positions were initialized to the positions of the final frame in the previous iteration. Here, neither energy-minimization nor an external force were applied during equilibration. Since these short simulations were insufficient for reliable estimation of~$\mathcal{L}$, longer validation simulations were performed every~$10$ iterations. These simulations followed the protocol employed for the multiphase reference simulations, but were initialized from the last frame of the corresponding optimization simulation. For reference, a multiphase optimization simulation required approximately~$2~\text{h}$ compute time on a single \textsc{Nvidia A5000} GPU, while a validation simulation required approximately~$18~\text{h}$.

For the optimization algorithms, the stepsize parameters~$r_\varepsilon$ and $\alpha$ were chosen to ensure comparable update magnitudes across different algorithms. A trust radius of~$r_\varepsilon=\sqrt{0.5}$ was used for all predictor-corrector~methods, while~$\alpha$ was set to~$\alpha=25$ in plain~GD. For Adam, a stepsize of~$\alpha=0.7$ was used together with parameters of~$\beta_1=0.7$, $\beta_2=0.9$ and~$\iota=10^{-8}$. The former parameters~$\beta_1, \beta_2$ were chosen to yield reasonably quick updates of the exponentially weighted averages used in the Adam algorithm. 

For all algorithms, a lower bound of~$\varepsilon_i \geq -\varsigma=-50\,k_\text{B} T$ was enforced by replacing all proposed~$\varepsilon_i$ exceeding this bound by~$-50\,k_\text{B} T$. This was done to prevent algorithms from proposing potentials in which particles could get trapped in the core of another particle. 

\subsection{\label{sec:Exp:UQprotocol}Observables \& Uncertainty Quantification}

For trajectory analysis, \textsc{MDanalysis 2.7.0},~\cite{MDanalysisCitation1, MDanalysisCitation2} and \textsc{MDtraj 1.9.9} were employed.~\cite{MDTrajCitation} Time series of all pair numbers in reference/validation simulations were visually inspected to ensure the absence of systematic drift. Based on the recommendations outlined by Grossfield~\textit{et al.},~\cite{BestPracticesUQMolSim} an approach combining block-averaging and bootstrapping was primarily used for uncertainty quantification. 

For each observable, integrated autocorrelation times~$\tau$ were estimated using \textsc{emcee 3.1.6}.~\cite{EmceeCitation} Here, the algorithm proposed by Sokal~\textit{et al.} was employed.~\cite{SokalInteAutoTime} From this, the value of an observable~$O$ was averaged over~$N_n$ blocks of size~$M=n\tau$ with~$n=5, 6, 7,\dots$ to obtain a set of approximately uncorrelated block averages~$\{\langle O_{i,n}\rangle\}_{i=1}^{N_n}$. These were then used to construct 95\%~confidence~intervals~(CIs)~$[O_{l,n}, O_{u,n}]$ with the bootstrapping technique implemented in \textsc{scipy 1.11.4} using the 'percentile' method.~\cite{SciPyCitation}

To mitigate the impact of correlation between blocks on the obtained CI estimates, the CI sizes~${|O_{s,n}|=|O_{u,n}-O_{l,n}|}$ were compared for different~$n$. Convergence of a CI estimate~$[O_{l,n^\ast}, O_{u,n^\ast}]$ was achieved when the CI size fell within 5\% of the average size of the three previous CI estimates. Unless otherwise noted, all quantities reported here refer to the mean~$O_{n^\ast}$ of the bootstrap distribution obtained from a converged~$n^\ast$ together with the 95\%~CI $[O_{l,n^\ast}, O_{u,n^\ast}]$. 

For composite quantities such as~$\mathcal{L}\left(\vecunder{z}\right)$, the same procedure was followed. However, the largest integrated autocorrelation time of the underlying observables (e.g.~$z_i$) was used as an estimate of~$\tau$ for constructing blocks. Meanwhile, the smallest considered~$n$ was~$n=1$, to reflect the rather conservative $\tau$~estimate. 

In all cases, the block size was increased until only ${N_n=5}$~blocks could be obtained from the trajectory. In rare cases where convergence was not observed before reaching~${N_n=5}$, the mean and CI of the~$n$ yielding the largest uncertainty span~$|O_{s,n}|$ will be reported and a note will be given in the text. 

\subsection{Numerical Methods}

HNC~solutions were obtained using a modified version of the \textsc{SunlightHNC 1.13} code.~\cite{SunlightHNC} Distribution functions and potentials were discretized on an equidistant grid of~$N_g=2^{15}$~points ranging from~$r_0=0~\text{nm}$ to~$r_{\text{max}}=300~\text{nm}$. 

When sampling~$g(r)$ from a MD simulation, an equidistant grid of~$N_g=1700$ points ranging from~$r_0=0~\text{nm}$ to $r_{\text{max}}=17~\text{nm}$ was used. This yielded similar values of~${\Delta r\approx0.00916~\text{nm}}$ and~${\Delta r=(r_{\text{max}}/N_g)=0.01~\text{nm}}$ for the HNC~and MD~grids, respectively. LDL derivatives and the LDL model were computed using the respective grid originating either from the HNC~solution or MD sampling. 

All integrals were calculated using Simpson's rule as implemented in the \textsc{simpson} function in \textsc{scipy 1.11.4}.~\cite{SciPyCitation} The discretization error was estimated by comparing results obtained from evaluating the integral in~$z[g]$ (eqn.~\ref{eqn:ConnectionNAB-gR}) with both Simpson's and the trapezoidal rule. The found relative mean absolute deviation between those results was typically less than~$10^{-5}$ and thus negligible compared to sampling uncertainties from MD results. Accordingly, discretization errors were generally neglected in this work.
\comment{, however, as will be outlined, the impact of~$N_g$ on the HNC~solutions was further investigated by comparing HNC~results for~$N_g=2^{15}$ and~$N_g=2^{16}$ in Section~\ref{sec:RnD_Opt_Singlephase}.}

In predictor-corrector~methods, the~$\mathcal{L}$ of predictor models was minimized using the Sequential Least Squares Programming (SLSQP) method implemented in the \textsc{minimize} function in \textsc{scipy~1.11.4}.~\cite{SciPyCitation} Here, the solution accuracy was controlled by setting~$\text{ftol}=10^{-10}$ and the trust radius was enforced through a constraint~on~$\vecunder{\varepsilon}$.

\section{\label{sec:Results} Results \& Discussion}

To evaluate the performance and robustness of the proposed optimization algorithms, tests were performed on two ten-component model systems. These systems were largely inspired by the tegument of human cytomegalovirus and designed to benchmark algorithmic performance in a biologically plausible setting. For both systems, reference pair numbers~$\vecunder{z}^\ast$ were obtained by sampling a set of reference interaction parameters~$\varepsilon_i^\ast$ from a normal distribution and performing multiple MD simulations using~$\vecunder{\varepsilon}^\ast$. The used simulation protocols are detailed in Appendix~\ref{chp:AppendixReferenceSimulations}.

The two systems were designed to be in two distinct thermodynamic regimes, one being homogeneous while the other formed a phase-separated mixture. Notes on the thermodynamic states of both systems can be found in Appendix~\ref{chp:AppendixReferenceSimulations}. For the homogeneous system, the volume was three times larger than the corresponding tegument volume, while the multiphase system used the approximate tegument volume. In the homogeneous system, the reference~$\vecunder{\varepsilon}^\ast$ was sampled from a normal distribution with mean~$\langle\varepsilon\rangle=-2\,k_\text{B} T$ and standard deviation~$\sigma_\varepsilon=1\,k_\text{B} T$, whereas in the multiphase system, $\langle\varepsilon\rangle=-3\,k_\text{B} T$~and~$\sigma_\varepsilon=3\,k_\text{B} T$ were employed. In all cases, the optimization algorithms were tested by fitting the reference~$\vecunder{z}^\ast$ starting from~$\vecunder{\varepsilon}= 0\,k_\text{B} T$ and performing $100$~optimization iterations.

\subsection{\label{sec:Results:Homogeneous} Homogeneous Systems}

The methods in Section~\ref{sec:Theory:OptAlgos} were derived specifically for multiphase systems. However, to demonstrate their validity, they will first be applied to the homogeneous ten-component reference system. In homogeneous systems, the HNC~approximation can be employed to evaluate~$\vecunder{\mathcal{G}}$ efficiently. Since HNC~solutions do not allow for the evaluation of the exact derivative, optimization algorithms are then limited to the trust-LDL algorithm as well as the gradient-based algorithms employing the LDL derivative approximation. 
To gauge the precision of the reference~$\vecunder{z}^\ast$, the uncertainty in $\mathcal{L}$ was estimated from the fluctuations observed during the reference simulations. This yielded an estimate for the ground-truth~$\mathcal{L}$ as~$\mathcal{L}\approx{4\times10^{-5}}$.

\begin{figure*}
    \centering
    \includegraphics[width=7in]{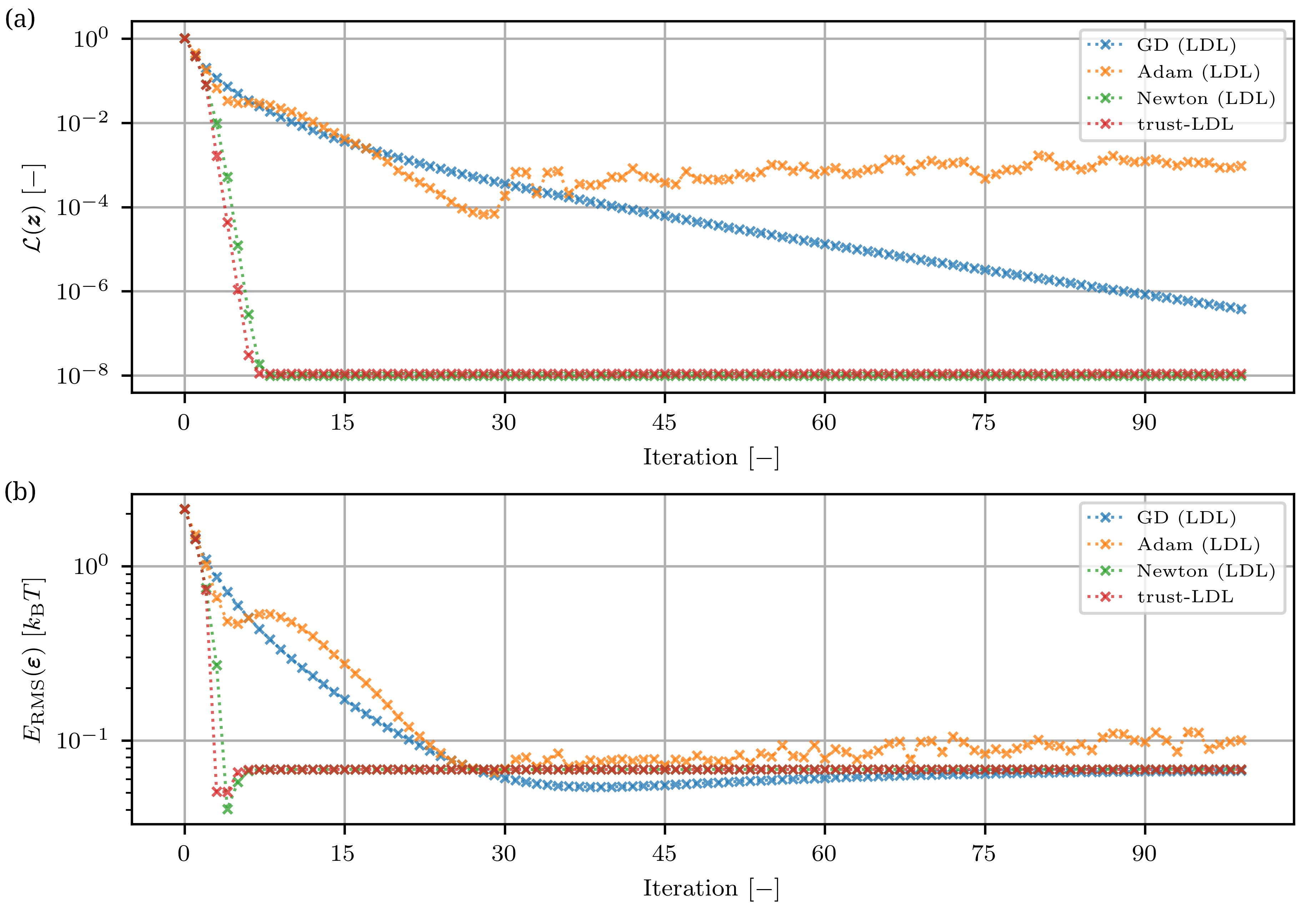}
    \caption{Learning curves for the optimization algorithms introduced in Section~\ref{sec:Theory:OptAlgos}, for values of (a)~$\mathcal{L}$, and (b)~$E_\text{RMS}$, during optimization with the four LDL algorithms using the HNC~equation to evaluate~$\vecunder{\mathcal{G}}$.
    \label{fig:RnDHomoHNC}}
\end{figure*}
\comment{
\begin{figure*}
    \centering
    \includegraphics{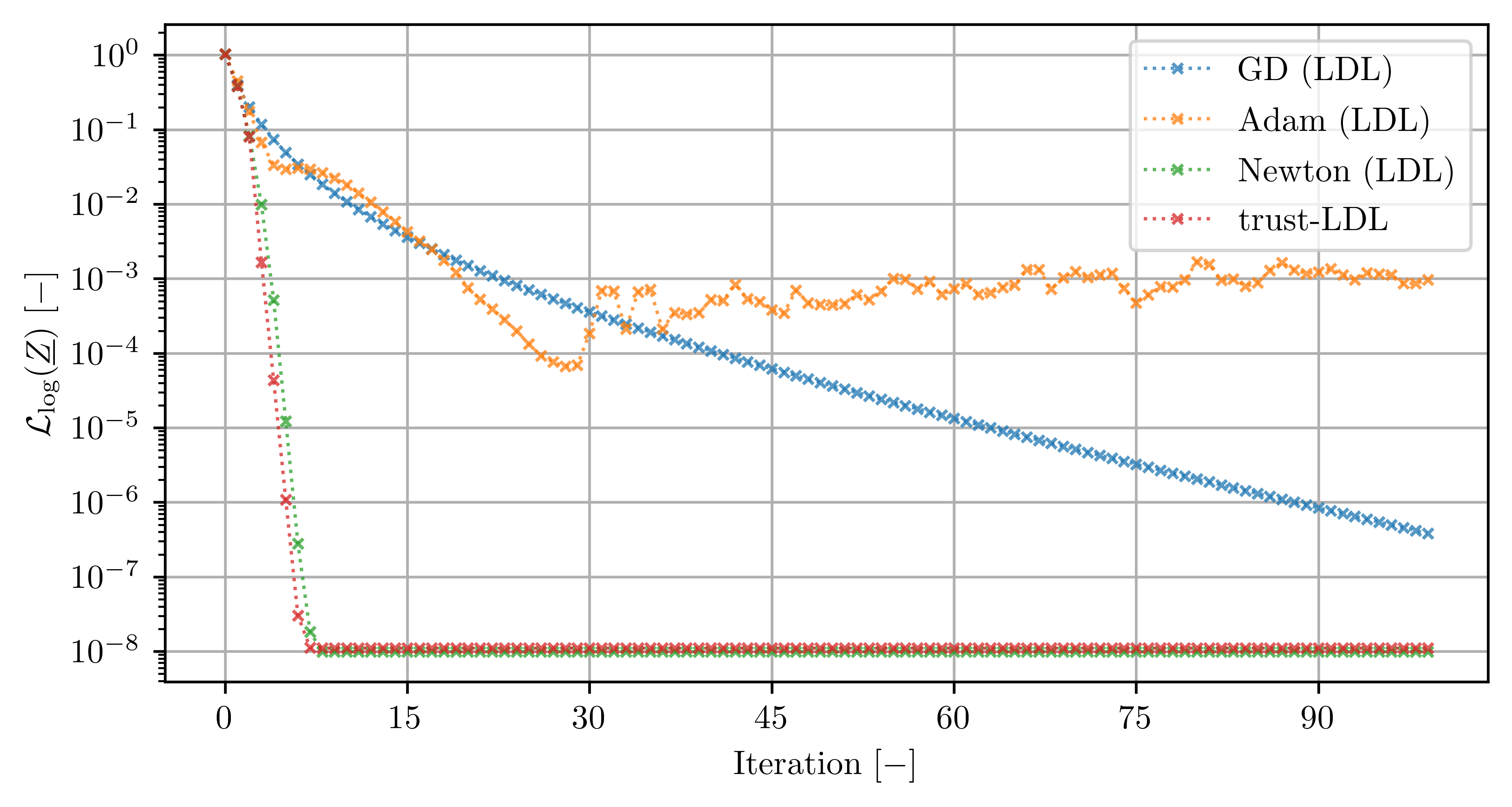}
    \caption{Value of~$\mathcal{L}$ during optimization with the four LDL algorithms outlined in Section~\ref{sec:Theory:OptAlgos} using the HNC~equation to evaluate~$\vecunder{\mathcal{G}}$.
    \label{fig:RnDHomoHNCLoss}}
\end{figure*}

\begin{figure*}
    \centering
    \includegraphics{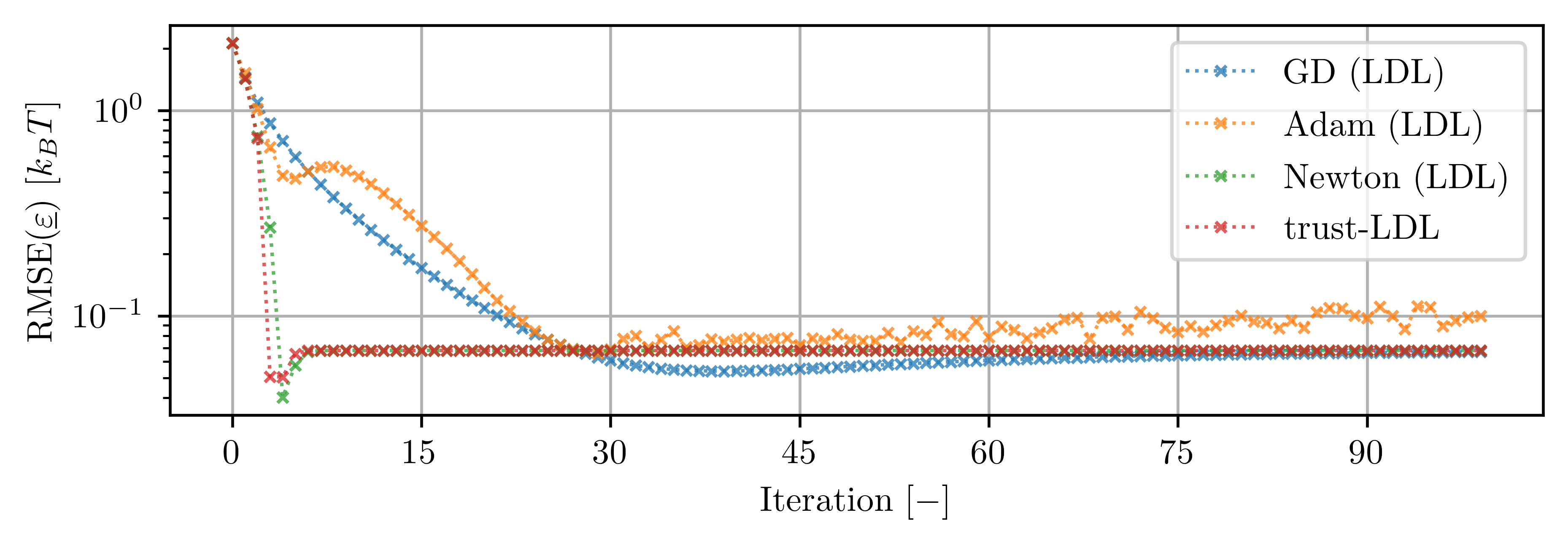}
    \caption{Value of~$E_\text{RMS}$ during optimization with the four LDL algorithms outlined in Section~\ref{sec:Theory:OptAlgos} using the HNC~equation to evaluate~$\vecunder{\mathcal{G}}$.
    \label{fig:RnDHomoHNCRMSE}}
\end{figure*}
}

Results of the optimization runs are depicted in Figure~\ref{fig:RnDHomoHNC}, measured using both~$\mathcal{L}$ and through the root mean squared error of the parameters of the interaction potentials,
\begin{equation}
    E_\text{RMS}\left(\vecunder{\varepsilon}\right)= \sqrt{\frac{1}{Q} \sum_{i=1}^{Q} \left(\varepsilon_i - \varepsilon_i^\ast\right)^2}
\label{eqn:EpsRMSE}
\end{equation}
compared to the reference~$\vecunder{\varepsilon}^\ast$. Both the trust-LDL and Newton~(LDL) algorithms exhibit rapid convergence, reducing the loss to~$\mathcal{L}\approx10^{-8}$ within only 8~iterations. Once this precision is reached, the convergence criterion of the predictor step minimization routine is satisfied, preventing any further changes. In contrast, the GD-based methods converge significantly slower. GD~(LDL) requires 100~iterations to reach~${\mathcal{L}\approx10^{-6}}$, while Adam~(LDL) stalls at~${\mathcal{L}\approx7\times10^{-5}}$ after approximately 30~iterations. Generally, all methods reach an accuracy that is comparable to the reference value of~$\mathcal{L}\approx4\times 10^{-5}$.

Considering the~$E_\text{RMS}$ depicted in Figure~\ref{fig:RnDHomoHNC}(b), a similar pattern can be observed regarding the convergence properties of the algorithms. Both predictor-corrector~methods converge rapidly within 8~iterations, while Adam~(LDL) stalls at~$E_\text{RMS}\approx\,0.07\,k_\text{B} T$ after approximately 30~iterations and GD~(LDL) reaches a minimum of~$E_\text{RMS}\approx\,0.05\,k_\text{B} T$ after approximately 40~iterations before showing higher~$E_\text{RMS}$ in later iterations. Notably, all algorithms that achieve~$\mathcal{L}\approx10^{-4}$ exhibit a similar pattern where further reduction in~$\mathcal{L}$ beyond this threshold leads to \textit{increasing}~$E_\text{RMS}$. This suggests systematic deviations between the HNC~evaluation and MD reference data.

All methods reach an accuracy of~$E_\text{RMS}\approx\,0.07\,k_\text{B} T$ at some point during optimization (Table~\ref{tab:RnDHomoHNCresults}). Here, both metrics are listed for the iteration with lowest~$\mathcal{L}$, which could be regarded as the best fit in absence of a ground truth $\vecunder{\varepsilon}^\ast$, i.e. in an experimental setting. At this iteration, all methods attain practically the same accuracy of~$E_\text{RMS}\lessapprox\,0.068\,k_\text{B} T$, while exhibiting different $\mathcal{L}$ values. 

These results clearly demonstrate the capability of all four algorithms to fit the reference data with sufficient accuracy. Especially noteworthy is the exceptional convergence of the predictor-corrector~methods, which achieve near-optimal accuracy within just $8$~iterations. This suggests that the LDL~approximation serves as an accurate predictor model in this relatively low-density system, which is then likely most effectively exploited by the predictor-corrector~methods. Thus, these approaches are very promising for solving the Inverse Pair Problem in homogeneous systems.

\begin{table}
\caption{Values~of~$\mathcal{L}$ and $E_\text{RMS}$ at the iteration of lowest~$\mathcal{L}$ for all four methods used in homogeneous systems.\label{tab:RnDHomoHNCresults}}
    \begin{tabular}{c|c|ccc}
        Method & Iteration & $\mathcal{L}~[-]$ &  $E_\text{RMS}~[k_\text{B} T]$\\\hline  
        \vphantom{$\left(10^{-7}\right)^2$}GD (LDL) & 99 & $3.82\times 10^{-7}$ & 0.0670\\[1pt]
        Adam (LDL) & 29 & $6.69\times 10^{-5}$ & 0.0677\\[1pt]
        Newton (LDL) & 8& $9.89\times 10^{-9}$ & 0.0680\\[1pt]
        trust-LDL & 8& $1.10\times 10^{-8}$ & 0.0680
    \end{tabular}
\end{table}
Regarding the accuracy achieved with each method, the residual deviations listed in table~\ref{tab:RnDHomoHNCresults} can be attributed to two factors. Firstly, the HNC~approximation introduces systematic errors and secondly, the MD simulation used to evaluate the reference pair numbers~$\vecunder{z}^\ast$ suffers from inherently limited sampling. This introduces some statistical fluctuations into the reference data, which correspond to a value of $\mathcal{L}\approx4\times 10^{-5}$. Two observations support the suggestion that the final accuracy is mainly limited by the precision of the optimization setup rather than algorithmic shortcomings. Firstly, the algorithms in Figure~\ref{fig:RnDHomoHNC} exhibit increasing~$E_\text{RMS}$ for smaller~$\mathcal{L}$ beyond~$\mathcal{L}\approx10^{-4}$ and secondly, there does not seem to be a clear relationship between the residual~$\mathcal{L}$ and~$E_\text{RMS}$ at the attained level of accuracy. This suggests that the accuracy of~$E_\text{RMS}\lessapprox\,0.068\,k_\text{B} T$ attained by all algorithms is likely due to systematic deviations inherent in the optimization setup rather than insufficient algorithmic accuracy.

Beyond verifying the validity of the chosen approaches, the results also demonstrate the computational efficiency of combining the algorithms with HNC~evaluation. The combined runtime for all four optimizations was approximately 2~hours on a single six-core CPU. Consequently, these results demonstrate that the proposed methods offer a highly efficient and promising approach for solving the Inverse Pair Problem in homogeneous fluids. 

\subsection{\label{sec:Results:InHomogeneous}Inhomogeneous Systems}
The HNC~approximation allows for efficient optimization in homogeneous fluids, however, multiphase systems require MD simulations to evaluate~$\vecunder{\mathcal{G}}$. While computationally expensive, MD simulations have the upside of yielding samples of the exact gradient, thus allowing for a comparison to employing the LDL~approximation during optimization. Since multiphase systems are difficult to equilibrate and sample, a two-tiered simulation scheme was employed. The evaluation of observables and gradients during the optimization steps was performed using short simulations of~$t_\text{prod}={500}~{\mu\text{s}}$. As these short simulations only yield noisy estimates of the loss function, a longer validation simulation ($t_\text{prod}={3}~{\text{ms}}$) was performed every $10$~iterations to accurately monitor the $\mathcal{L}$~convergence. As noted in Appendix~\ref{chp:AppendixReferenceSimulations}, the intrinsic fluctuations in the multiphase~$\vecunder{z}^\ast$ correspond to a reference accuracy of~$\mathcal{L}\approx{0.01}$.

\begin{figure*}
    \centering
    \includegraphics[width=7in]{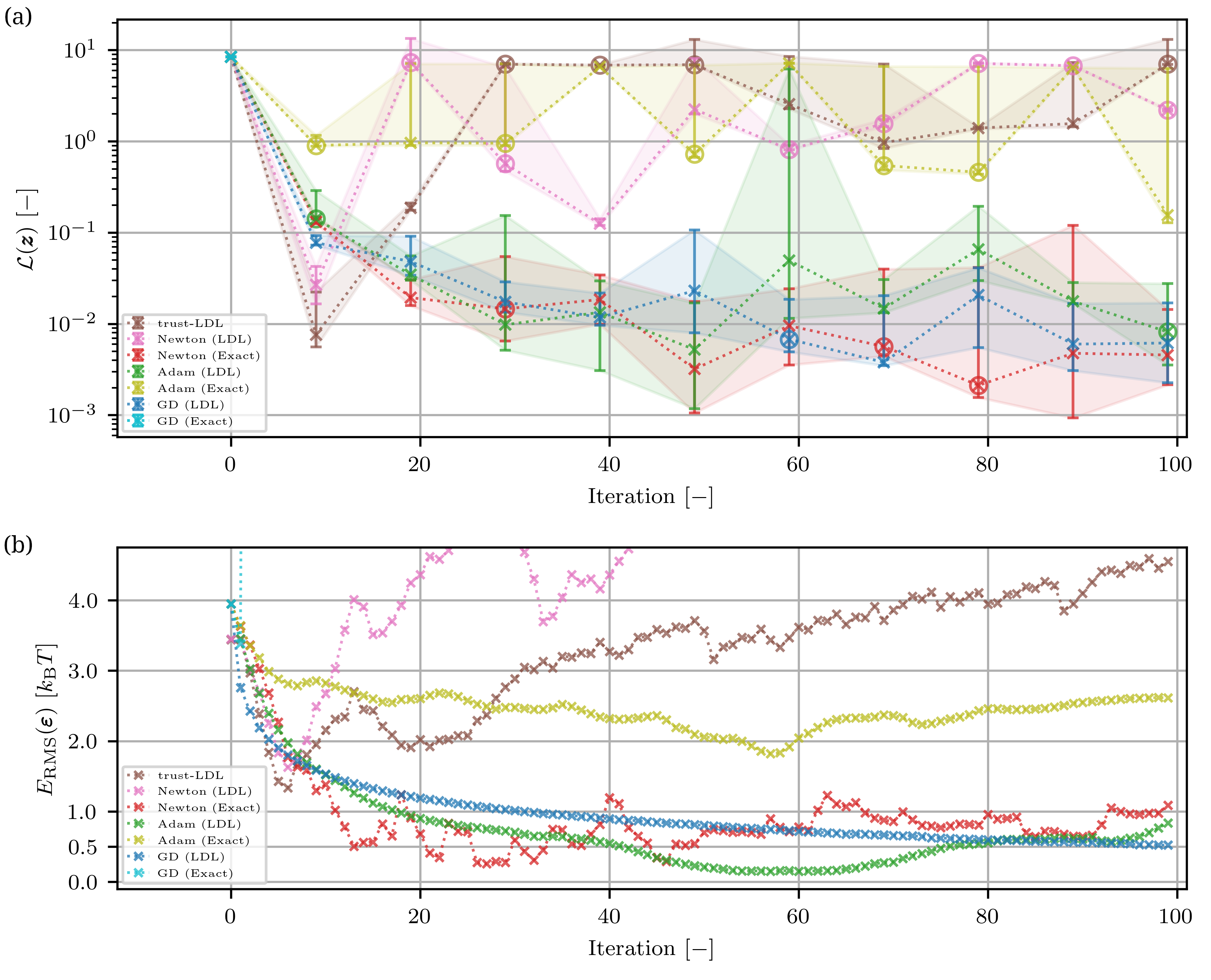}
    \caption{Similar to Fig.~\ref{fig:RnDHomoHNC} for optimization in the multiphase system, employing MD~simulations for the evaluation of~$\vecunder{\mathcal{G}}$. Simulations for which the uncertainty quantification protocol (see Section~\ref{sec:Exp:UQprotocol}) did not converge are marked by circles. (b) Value of the Parameter~$E_\text{RMS}$ during the same optimization run. For the full range of $E_\text{RMS}$~values please see Figure~\ref{fig:RnDMultiRMSEFull}. 
    \label{fig:RnDMultiResults}}
\end{figure*}

\begin{figure*}
    \centering
    \includegraphics[width=7in]{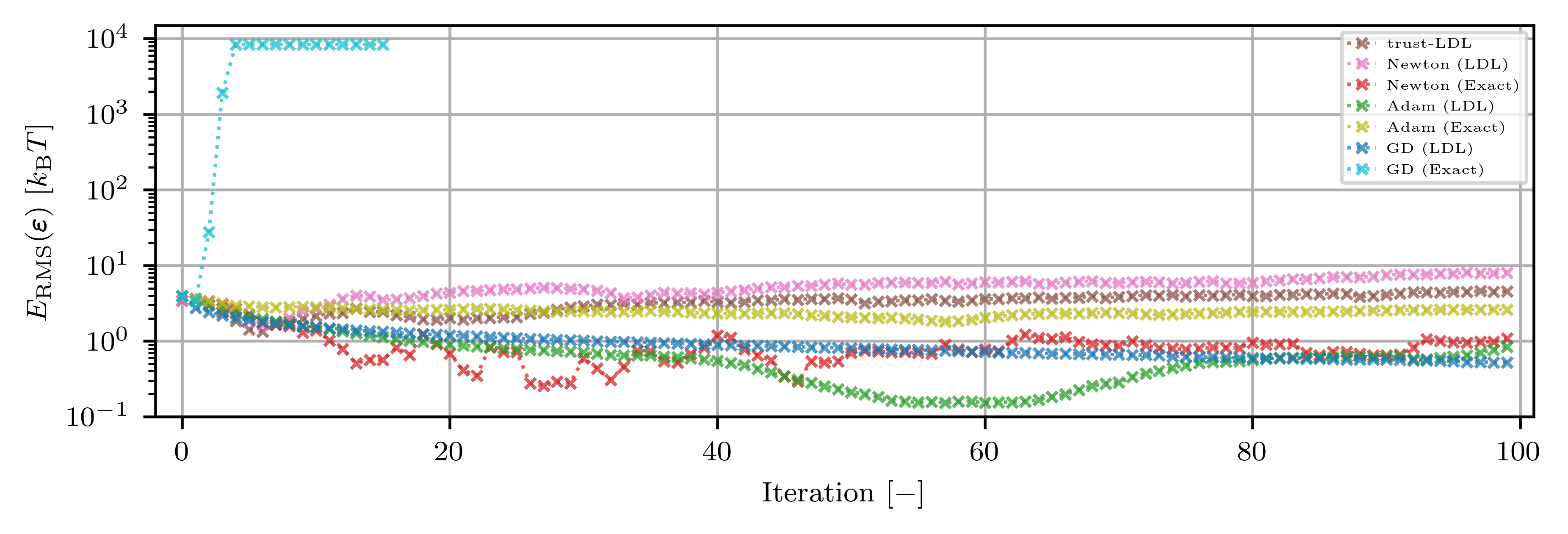}
    \caption{Value of~$E_\text{RMS}$ during optimization of the phase-separated system using the algorithms outlined in Section~\ref{sec:Theory:OptAlgos} employing MD~Simulations to evaluate~$\vecunder{\mathcal{G}}$.
    \label{fig:RnDMultiRMSEFull}}
\end{figure*}

The optimization performance varies significantly across the seven tested algorithms, as illustrated by the obtained validation~$\mathcal{L}$ which is displayed together with the parameter~$E_\text{RMS}$ in Figure~\ref{fig:RnDMultiResults}. For better legibility, the~$E_\text{RMS}$ is only shown partially in Figure~\ref{fig:RnDMultiResults}, the full range of $E_\text{RMS}$~values is displayed in Figure~\ref{fig:RnDMultiRMSEFull}. Based on the observed results, the methods can be broadly grouped into three categories. Firstly, GD~(LDL), Adam~(LDL) and Newton~(Exact) were successful in reproducing the reference parameters and minimizing~$\mathcal{L}$. In contrast, Newton~(LDL) and trust-LDL achieved similar precision in~$\mathcal{L}$, but failed to ultimately produce the correct set of parameters. Lastly, GD~(Exact) and Adam~(Exact) simply failed to reach adequate accuracy.

For the three successful algorithms, all three reached the reference accuracy of~$\mathcal{L}\approx{0.01}$ within approximately $40$~iterations. Subsequently, only slight further improvement was observed for GD~(LDL) and Newton~(Exact), while Adam~(LDL) did not show any systematic trend. Despite similar final values of~$\mathcal{L}\lessapprox0.01$, the parameter~$E_\text{RMS}$ displayed in Figure~\ref{fig:RnDMultiResults}(b) reveals notable differences. While all three algorithms attain~$E_\text{RMS}\approx0.5\,k_\text{B} T$, Newton~(Exact) and Adam~(LDL) temporarily achieve even higher precision of~$E_\text{RMS}\approx0.25\,k_\text{B} T$ and~$E_\text{RMS}\approx0.15\,k_\text{B} T$ around iterations~$30$ and~$60$, respectively, which is not clearly reflected in the~$\mathcal{L}$ values. In later iterations, both methods display a slight decrease in accuracy, mirroring the behavior seen in homogeneous systems. Conversely, GD~(LDL) exhibited a slower but more monotonic decrease~in~$E_\text{RMS}$, maintaining steady improvement throughout all iterations while never attaining the same accuracy as the other two methods. 

Lastly, it should be noted that the uncertainty quantification protocol outlined in Section~\ref{sec:Exp:UQprotocol} did not converge for all validation simulations, such that some of the obtained confidence intervals might be underestimated. In Figure~\ref{fig:RnDMultiResults}(a), this is noted by circular markers. This issue is due to long-lived fluctuations inherent to multiphase interfaces, which persisted across different initialization protocols and simulation parameters. However, we also did not find considerable disagreement in the obtained~$\mathcal{L}$ values for different simulation protocols. So given that the overarching conclusions are robust to this issue, the existing validation results were deemed sufficient for this comparative study.

Considering the results of the algorithms displayed in Figure~\ref{fig:RnDMultiResults}(a), only Newton~(LDL) and trust-LDL attained~$\mathcal{L}\,<\,0.1$. Most striking is the behavior of the GD~(Exact)~$E_\text{RMS}$ displayed in Figure~\ref{fig:RnDMultiRMSEFull}, where the full $E_\text{RMS}$~values are shown. GD~(Exact) proved unstable, rapidly proposing non-physical parameter sets that eventually led to simulation crashes. For this reason $E_\text{RMS}$~data is absent beyond iteration $15$ and no validation simulation could be conducted for iteration $9$. However, it should be noted that the data shown after iteration $2$ is also unreliable since the (fixed) integration timestep was not sufficient for performing valid simulations with the proposed~$\vecunder{\varepsilon}$. Adam~(Exact) avoided such catastrophic divergence, however it also failed to achieve significant improvements, with~$\mathcal{L}$ and $E_\text{RMS}$ stalling far above the benchmark accuracy. Conversely, the LDL-based predictor-corrector~methods (Newton~(LDL) and trust-LDL) initially exhibited very strong performance, with trust-LDL achieving~$\mathcal{L}\approx0.01$ in iteration~$9$, and attaining the highest precision of all algorithms within the first $5$~iterations. However, beyond this initial success, i.e. after iteration~$6$, both methods diverged sharply proposing significantly less accurate parameters during the remainder of optimization.

\begin{table}
\caption{\label{tab:RnDMultiMDOpt} Values of~$\mathcal{L}$ and~$E_\text{RMS}$ at the iteration of lowest~$\mathcal{L}$ for all seven methods detailed in Section~\ref{sec:Theory:OptAlgos}. It should be noted that the uncertainty quantification protocol in Section~\ref{sec:Exp:UQprotocol} did not converge for the Newton~(Exact) result.}
    \centering
    \begin{tabular}{c|c|ccc}
        Method & Iteration & $\mathcal{L}~[-]$ &  $E_\text{RMS}~[k_\text{B} T]$\\ \hline
        \vphantom{$\left(8.4432^{+22}_{-27}\right)^a$}GD (Exact) ($\vecunder{\varepsilon}=\vecunder{0}\,k_\text{B} T$)  & 0  & $8.4432^{+22}_{-27}$ & $3.95$\\[2pt]
        Adam (Exact)  & 99 & $0.2^{+6.2}_{-0.1}$ & $2.62$\\[2pt]
        Newton (LDL)  & 9  & $0.027^{+16}_{-11}$ & $2.49$\\[2pt]
        trust-LDL     & 9  & $0.008^{+15}_{-02}$ & $1.95$\\[2pt]
        Newton (Exact)& 79 & $0.002^{+40}_{-01}$ & $0.808$\\[2pt] 
        GD (LDL)      & 69 & $0.004^{+17}_{-01}$ & $0.668$\\[2pt]
        Adam (LDL)    & 49 & $0.005^{+12}_{-05}$ & $0.232$
    \end{tabular}
\end{table}

In accordance with the results presented for homogeneous systems, the results for the iteration with lowest~$\mathcal{L}$ are listed in Table~\ref{tab:RnDMultiMDOpt}. Here, the obtained results are largely consistent with the previously discussed findings. GD~(Exact) failed to yield any improvements while Adam~(Exact) displayed only marginal advances beyond the initial parameter set. The two LDL predictor-corrector~schemes yielded notably improved~$\mathcal{L}$ initially, even reaching the benchmark value of~$\mathcal{L}\approx0.01$. However, particularly the attained~$E_\text{RMS}$ values are significantly larger than those achieved by the three most accurate algorithms. These displayed similar final values~of~$\mathcal{L}\approx0.004$ and attained $E_\text{RMS}$~values of~$E_\text{RMS}=0.808\,k_\text{B} T$, $E_\text{RMS}=0.668\,k_\text{B} T$ and $E_\text{RMS}=0.232\,k_\text{B} T$ for Newton~(Exact), GD~(LDL) and Adam~(LDL), respectively.

Overall, these results demonstrate that despite the complexity of phase-separated systems, three of the proposed optimization strategies (GD~(LDL), Adam~(LDL), and Newton~(Exact)) are capable of reproducing the correct parameters. Beyond these quantitative metrics, visual inspection of the optimization simulations revealed that the characteristic three-phase topology of the reference system was correctly reproduced even for relatively crude fits. For instance, this structural agreement was visible as early as the second or third iteration in the short optimization simulations of the LDL-based predictor-corrector~methods. This suggests that even relatively inaccurate solutions of the Inverse Pair Problem can capture the essential qualitative structure of the system. Thus demonstrating that the presented framework is relatively robust at resolving the essential information in complex mixtures, making it promising for structural discovery in complex biological systems where qualitative phase behavior is often the primary interest.

\paragraph{Interpretation of Algorithmic Performance}

The distinct performance hierarchy observed across the seven algorithms warrants a detailed discussion of the underlying derivative models and optimization strategies. Regarding the gradient-based approaches employing the exact derivative, the success of the Newton~(Exact) algorithm demonstrates that the MD simulations provided sufficiently accurate sampling of the Jacobian in order to perform optimization. However, the divergence of the GD~(Exact) algorithm suggests that the exact gradient can become exceedingly steep in multiphase systems. This is consistent with auxiliary tests on phase-separated systems, where steep gradients were observed, which may limit the admissible step size in plain GD.

In contrast, the success of GD~(LDL) and Adam~(LDL) can be attributed to the inherent mathematical properties of the LDL derivative approximation. It can be shown that the entries of the LDL gradient are naturally bounded. Specifically, for the chosen potential~(eqn.~\ref{eqn:ApproxPMF}) it holds that~${0\geq\partial_{\beta \varepsilon_i} u_i(r)\geq-1}$, leading to the LDL derivative of the pair numbers,
\begin{equation}
    \frac{\partial z_i}{\partial \beta \varepsilon_i}=-\frac{N_{a_i}N_{b_i}} {V (1+ \delta_{a_i b_i})} \int_0^{r_M} 4 \pi r^2 g_i(r) \frac{\partial u_i(r)}{\partial \beta \varepsilon_i} dr\ ,
\end{equation}
to follow
\begin{equation}
     0\leq \frac{\partial z_i}{\partial \beta \varepsilon_i}\leq z_i \ .
\end{equation}
Thus, the magnitude of the LDL gradient is effectively regularized by the employed approximation. This prevents the erratic updates observed with the exact derivative, making the LDL~approximation a robust choice for gradient-based optimization.

The failure of the LDL-based predictor-corrector~algorithms (Newton~(LDL) and Trust-LDL) in later iterations remains a peculiar result, especially given their initial rapid progress. The observed initial performance could be very misleading in an experimental setting, as the~$\mathcal{L}$ attained in iteration~$9$ might give a (false) impression of convergence. While this indicates that the inverse problem may be ill-conditioned for obtaining the ground-truth parameters, the observation that the qualitative properties such as the correct phase structure can be resolved at this relatively inaccurate~$E_\text{RMS}$, ameliorates this concern. Considering the behavior of the more well-behaved algorithms, the obtained results indicate that further iterations after achieving threshold accuracy are warranted in order to ensure reliable results for all algorithms. 

The initial progress, combined with the subsequent degradation in accuracy, indicates that the LDL~approximation alone might be misleading in multiphase systems. While the GD-based methods relying on the same LDL~approximation successfully obtain the reference parameters, the predictor-corrector~methods are likely less stable, as they rely more heavily on the accuracy of the employed predictor model. This can be seen in Figure~\ref{fig:RnDHomoHNC}(b) and~\ref{fig:RnDMultiResults}(b) where predictor-corrector~methods exhibit larger iteration-to-iteration variations, while the GD-based methods appear to yield smaller and more well-behaved updates. Thus, predictor-corrector~methods seem to be somewhat susceptible to performing large, potentially misleading parameter updates and thereby more sensitive to the underlying predictor model. While more elaborate algorithms such as line-search methods might mitigate such problems, they would likely also slow down optimization, making these approaches ultimately less competitive than the more stable GD-based counterparts.

The robustness of the Newton~(Exact) algorithm is also noteworthy. While it was observed that the exact gradient can become very steep, Newton~(Exact) appears relatively well-behaved. One possible explanation involves the interplay between the predictor model and the trust-radius constraint. In single-component systems, the coordination number~$z(\varepsilon)$ must remain bounded since it has to be nonnegative and there must be a maximum coordination number, as the omnipresent repulsive branch of the potential will prevent total collapse of neighboring particles. For steep gradients, the curvature of~$z(\varepsilon)$ must then cause the (unbounded) linear predictor model to underestimate the $\varepsilon$~update required to reach a target~$z^\ast$. This underestimation could act as an implicit regularization in regions with steep gradients. Conversely, for shallow gradients where the linear model might overestimate required updates, the trust-radius explicitly limits step sizes. Together, these mechanisms may contribute to the observed stability, though further investigation would be necessary to confirm this hypothesis. 

Similar to the discussion outlined for homogeneous systems, the precision of the reference simulations was estimated to lead to a nonzero value of~$\mathcal{L}\lessapprox0.01$ for the ground truth~$\vecunder{\varepsilon}^\ast$. Thus, the attained accuracy of~$\mathcal{L}\lessapprox0.004$ confirms that the optimization has reached the benchmark precision permitted by the chosen simulation parameters. This is further supported by the fact that algorithmic progress stalls for all algorithms at this level, and that there does not seem to be a clear relationship between~$E_\text{RMS}$ and~$\mathcal{L}$ in Figure~\ref{fig:RnDMultiResults} beyond~$\mathcal{L}\lessapprox0.01$. Thus, the residual~$E_\text{RMS}$ values are likely due to the chosen simulation parameters rather than algorithmic limitations. 

Lastly it should be noted that, while the results presented in table~\ref{tab:RnDMultiMDOpt} may suggest that Adam~(LDL) is generally more accurate and faster converging than GD~(LDL) and Newton~(Exact), we did not make this observation when testing different validation simulation protocols. Instead, we found that different simulation protocols yielded consistent $\mathcal{L}$ values for all validation simulations beyond iteration~$40$. Thereby, $\mathcal{L}$~fluctuations led to different iterations having the lowest~$\mathcal{L}$, with typical final values of $E_\text{RMS}\approx0.5\,k_\text{B} T$~for all three algorithms. 

Overall, these results demonstrate that the proposed methods offer a powerful framework for solving the Inverse Pair Problem in multiphase systems and provide a significant step forward for fitting interaction parameters to experimental observables. This is particularly valuable for applications in structural biology, such as the refinement of CG potentials based on XL-MS results, where the ability to resolve the underlying phase structure of the system is paramount.

\section{\label{sec:Conclusions} Conclusions}
This work established methods for systematically determining interaction potentials from coarse-grained experimental observables in homogeneous and phase-separated systems. By establishing a theoretical framework connecting cross-linking mass spectrometry data with interaction parameters, we enable the interpretation of XL-MS experiments through quantitative coarse-grained models, providing a route to computational modeling from XL-MS results.

We developed and rigorously tested seven optimization algorithms for fitting experimental observables. Here, two derivative approximations were employed to facilitate optimization, the exact gradient sampled from MD simulations, and a low-density limit approximation computed from the radial distribution function. Four optimization algorithms were evaluated in both homogeneous and phase-separated systems. These algorithms can be grouped into gradient descent (GD) algorithms (plain~GD and Adam) and predictor-corrector~methods (Newton's method and the trust-LDL algorithm).

In homogeneous systems, all algorithms demonstrated exceptional efficiency when combined with the hypernetted~chain~approximation to evaluate radial distribution functions. Here, the predictor-corrector~methods rapidly achieve highly accurate results. Crucially, we further demonstrate that the established algorithms can also solve the problem in the challenging regime of phase-separated mixtures. Here, three methods (Adam~(LDL), GD~(LDL), and Newton~(Exact)) showed robust results and successfully recovered the reference parameters, with Adam~(LDL) and Newton~(Exact) appearing especially capable. Notably, even approximate solutions correctly captured the qualitative three-phase topology, demonstrating that the framework can reliably resolve essential structural features. Thus, these results establish a practical foundation for parameterizing coarse-grained models in potentially phase-separated mixtures and offer a promising route for gaining new insights into phase-separated systems.

\begin{acknowledgments}
The authors are especially thankful to Fan Liu (Leibniz-Forschungsinstitut f\"{u}r Molekulare Pharmakologie) and Boris Bogdanow (Charit\'{e}--Universit\"{a}tsmedizin Berlin) for providing XL-MS data and insights. This research has been funded by Deutsche Forschungsgemeinschaft (DFG) through grant SFB 1114.
\end{acknowledgments}

\section*{Data Availability Statement}
The data that support the findings of this study are openly available from the ftp server ftp.imp.fu-berlin.de/msadeghi/xlms\_interaction\_inference, with the underlying code available at https://git.imp.fu-berlin.de/boerriev00/pubircpkg.


\appendix

\section{\label{chp:AppendixReferenceSimulations} Reference Simulations}
\paragraph{Simulation Protocol}
The systems used in Section~\ref{sec:Results:InHomogeneous} and~\ref{sec:Results:Homogeneous} were constructed to test the optimization methods outlined in Section~\ref{sec:Theory:OptAlgos} under conditions similar to what would be expected from biological systems. Accordingly, the particle types used in both systems were largely inspired by the ten most abundant proteins found in the tegument of human cytomegalovirus. For these proteins, accurate experimental expression numbers were available.~\cite{HCMmodelMohsen} The employed particle numbers are listed in Table~\ref{tab:ProteinData} together with the employed particle~radii~$r^m_a$, which were estimated as the radius of gyration of the \textsc{AlphaFold2} model of the corresponding folded protein.~\cite{AlphaFold2}

\begin{table*}[t]
    \caption{\label{tab:ProteinData}Particle number~$N_a$ and radius~$r_{a}^m$ of the different particles used in the ten-component reference systems.}
    \centering
    \begin{tabular}{l|llllllllll}
        Name & UL132 & UL88 & UL48 & UL47 & UL94 & UL35 & UL100 & UL25 & UL82 & UL83 \\ \hline
        $r_{a}^m$ [\AA] & 37.2 & 23.3 & 58.3 & 40.3 & 22.7 & 36.1 & 28.2 & 37.1 & 31.6 & 30.7 \\ 
        $N_a$ & 102 & 117 & 120 & 127 & 349 & 446 & 553 & 567 & 2030 & 2094   
    \end{tabular}
\end{table*}

The homogeneous and multiphase reference systems differ primarily in their simulation volumes and force field parameters. These details are outlined in Section~\ref{sec:Exp:Systems} as well as in the beginning of Section~\ref{sec:Results}. To obtain accurate~$\vecunder{z}^\ast$ estimates, three distinct simulations were conducted for each system, employing the parameters listed in Section~\ref{sec:Exp:SimParams}.

\begin{figure*}
    \centering
    \includegraphics[height=7.8in]{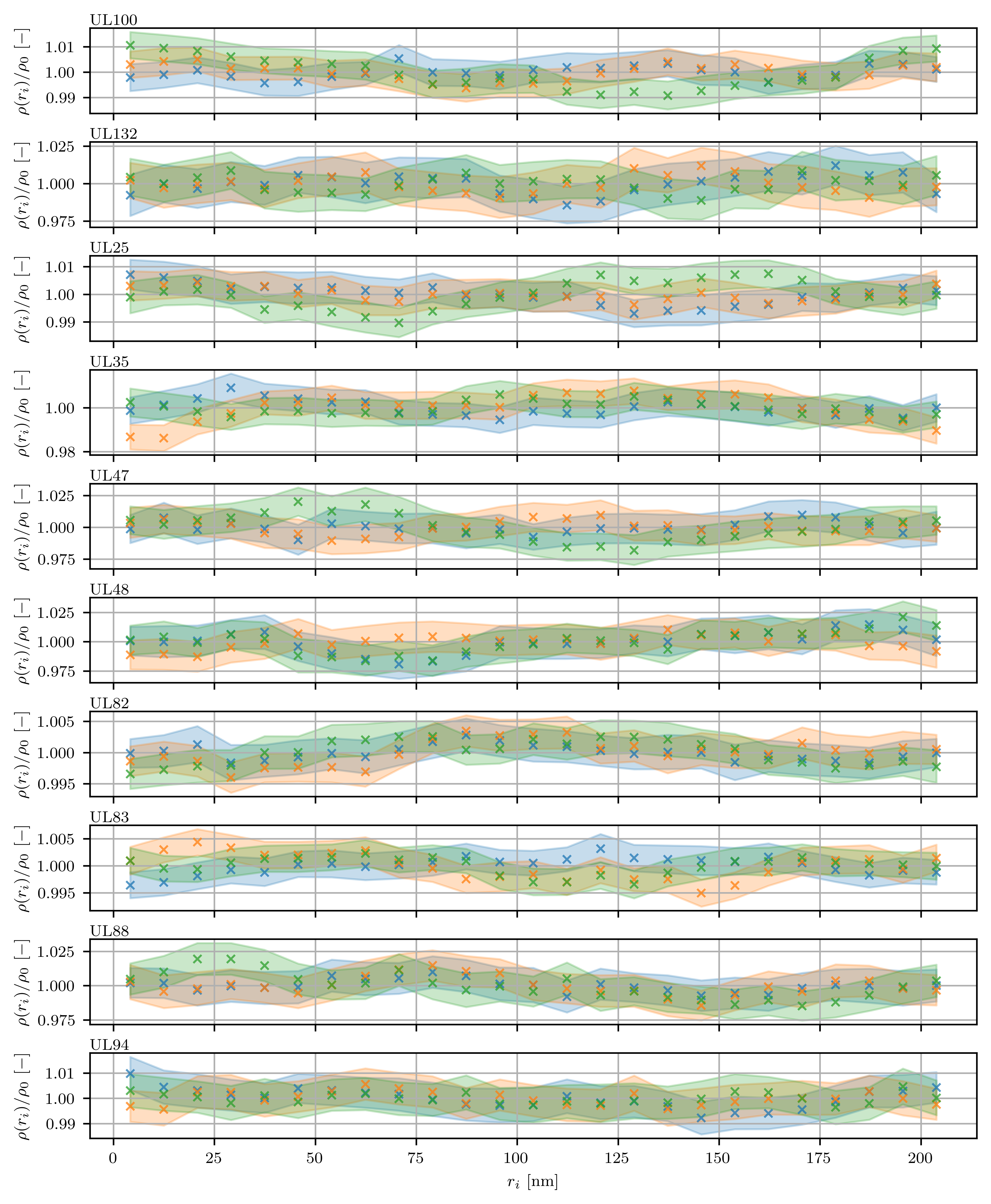}
    \caption{\label{fig:LinRhoSinglePhase}Local density profiles in the homogeneous reference system. The species type is specified in the top-left corner of each plot, while the colors denote the axes along with the system was split. The x, y and z axes are depicted in blue, orange and green, respectively. The shaded regions indicate 95\% confidence intervals and the profiles were normalized w.r.t. the respective total number density~$\rho_0=N_a/V$.}
\end{figure*}

\begin{figure*}
    \centering
    \includegraphics[height=7.8in]{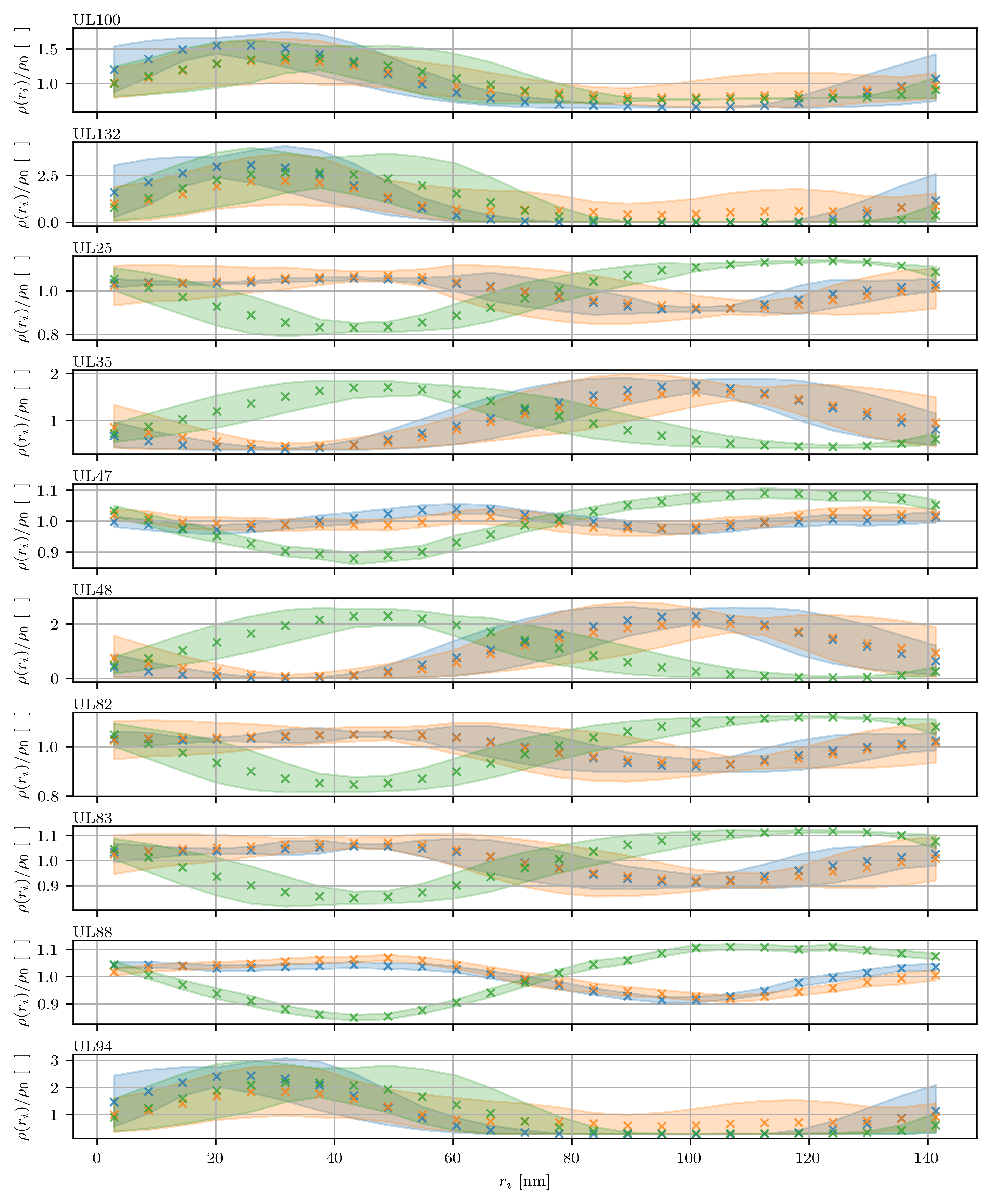}
    \caption{\label{fig:LinRhoMultiPhase}Local density profiles in the multiphase reference system. The species type is specified in the top-left corner of each plot, while the colors denote the axes along with the system was split. The x, y and z axes are depicted in blue, orange and green, respectively. The shaded regions indicate 95\% confidence intervals and the profiles were normalized w.r.t. the respective total number density~$\rho_0=N_a/V$.}
\end{figure*}

\paragraph{State Point Analysis}
The multiphase reference system was constructed to test the capabilities of each algorithm to cope with systems containing multiple phases, whereas the homogeneous system was constructed to test algorithms limited to uniform mixtures.
In the multiphase reference simulations, visual inspection revealed three spatially segregated regions with distinct compositions, whereas the homogeneous reference systems appeared uniform. This is corroborated by Figure~\ref{fig:LinRhoSinglePhase} and Figure~\ref{fig:LinRhoMultiPhase} which depict density profiles for each particle type in one representative reference simulation. To obtain these density profiles, the cubic simulation box was dissected into 25 equal-width slabs along each of the three spatial axes~$r_i$. The mean local number density~$\rho(r_i)$ for each slab was then recorded and is displayed in the figures. 

The density profiles displayed in Figure~\ref{fig:LinRhoMultiPhase} confirm the heterogeneity of the multiphase system. As shown, all densities show substantial deviations from the mean density~$\rho_0$. Further, particles UL94, UL100, and UL132 display strongly correlated density peaks around~$r_i=30~\text{nm}$ for all three axes, suggesting co-localization within a shared condensed phase. Similarly, UL35 and UL48 exhibit correlated density maxima at different spatial locations ($r_x,\,r_y\approx100~\text{nm}$,~$r_z=50~\text{nm}$), suggesting a second distinct phase. The remaining particle types show intermediate behavior, likely forming a third, less dense phase. In the homogeneous system, such spatial correlations are absent, which is shown in Figure~\ref{fig:LinRhoSinglePhase}. Here, all densities fluctuate randomly around~$\rho_0$ with deviations remaining within 3\%~of~$\rho_0$, indicating uniform mixing.

\begin{figure*}
    \centering
    \includegraphics[width=\textwidth]{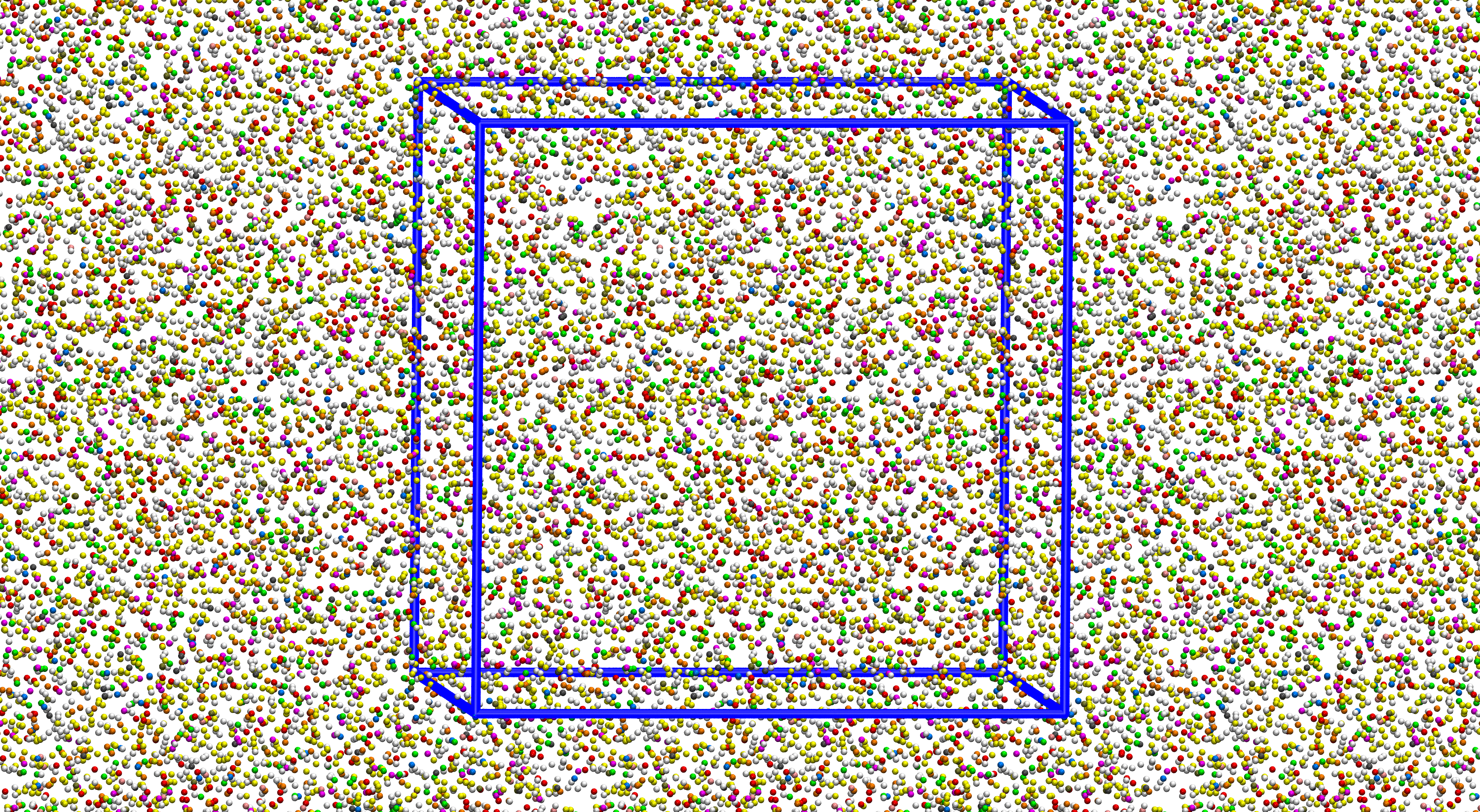}
    \caption{\label{fig:SnapHomogeneous}Snapshot from one of the reference simulations of the homogeneous reference system. The simulation box is outlined in blue and surrounding copies are periodic images. The species are colored as follows: UL100: red; UL132: grey; UL25: orange; UL35: green; UL47: olive drab / dark yellow; UL48: blue; UL82: yellow; UL83: white; UL88: light pink; UL94: violet.}
\end{figure*}

\begin{figure*}
    \centering
    \includegraphics[width=\textwidth]{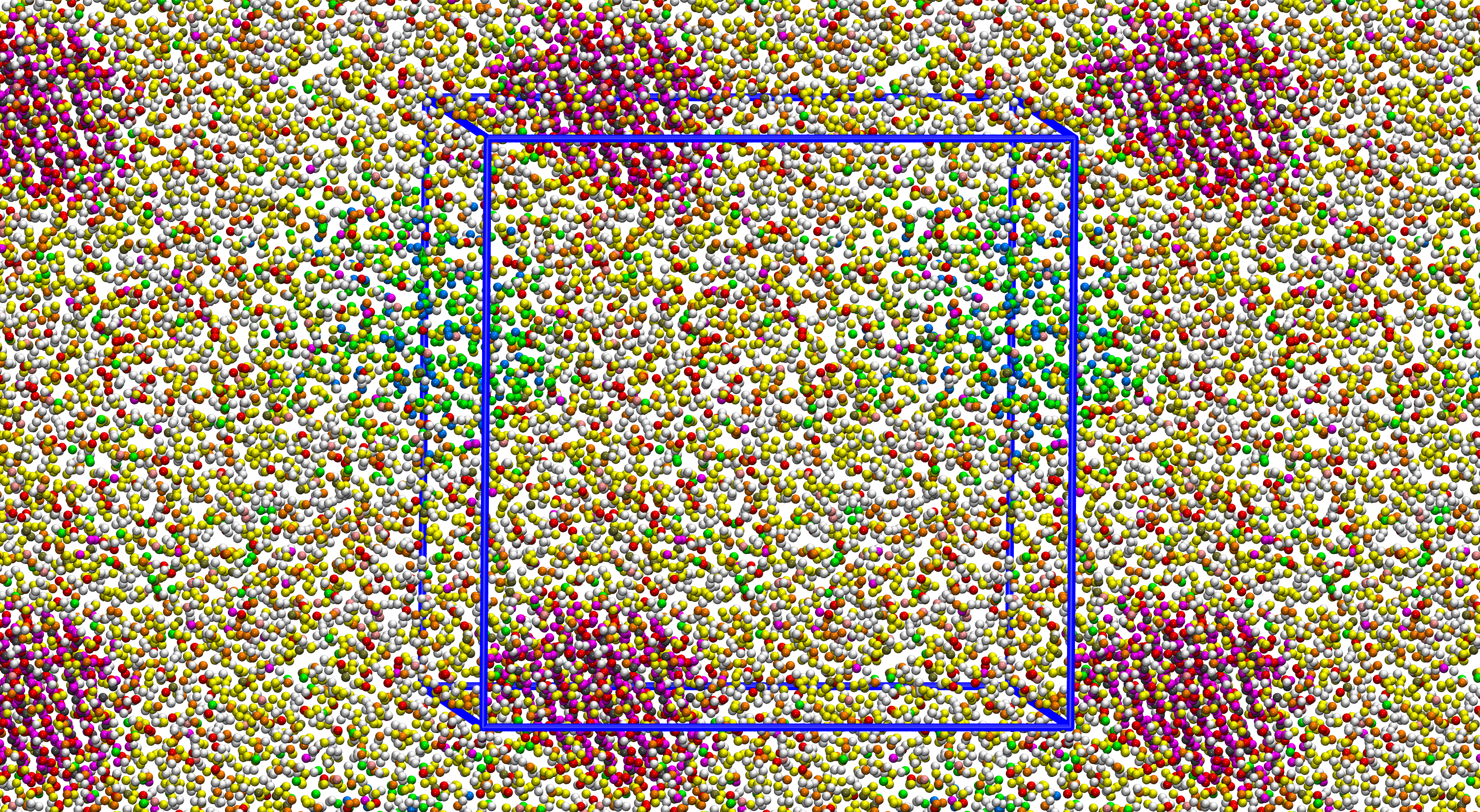}
    \caption{\label{fig:SnapMulti}Snapshot from one of the reference simulations of the multiphase reference system. The simulation box is outlined in blue and surrounding copies are periodic images. The coloring follows the convention in Figure~\ref{fig:SnapHomogeneous}.}
\end{figure*}

These findings are also consistent with the observations from visual inspections. Snapshots of exemplary simulations are displayed in Figure~\ref{fig:SnapHomogeneous} and Figure~\ref{fig:SnapMulti}, where the phase separated nature of the multiphase system is clearly visible, while the homogeneous system displays a uniform distribution of all particle types. These snapshots further corroborate the conclusions from Figure~\ref{fig:LinRhoMultiPhase}, as the phase formed by UL94, UL100 and UL132 appears in violet, red and grey, while the phase consisting of UL35 and UL48 is rendered in green and blue. 

\begin{figure*}
    \centering
    \includegraphics[height=7.8in]{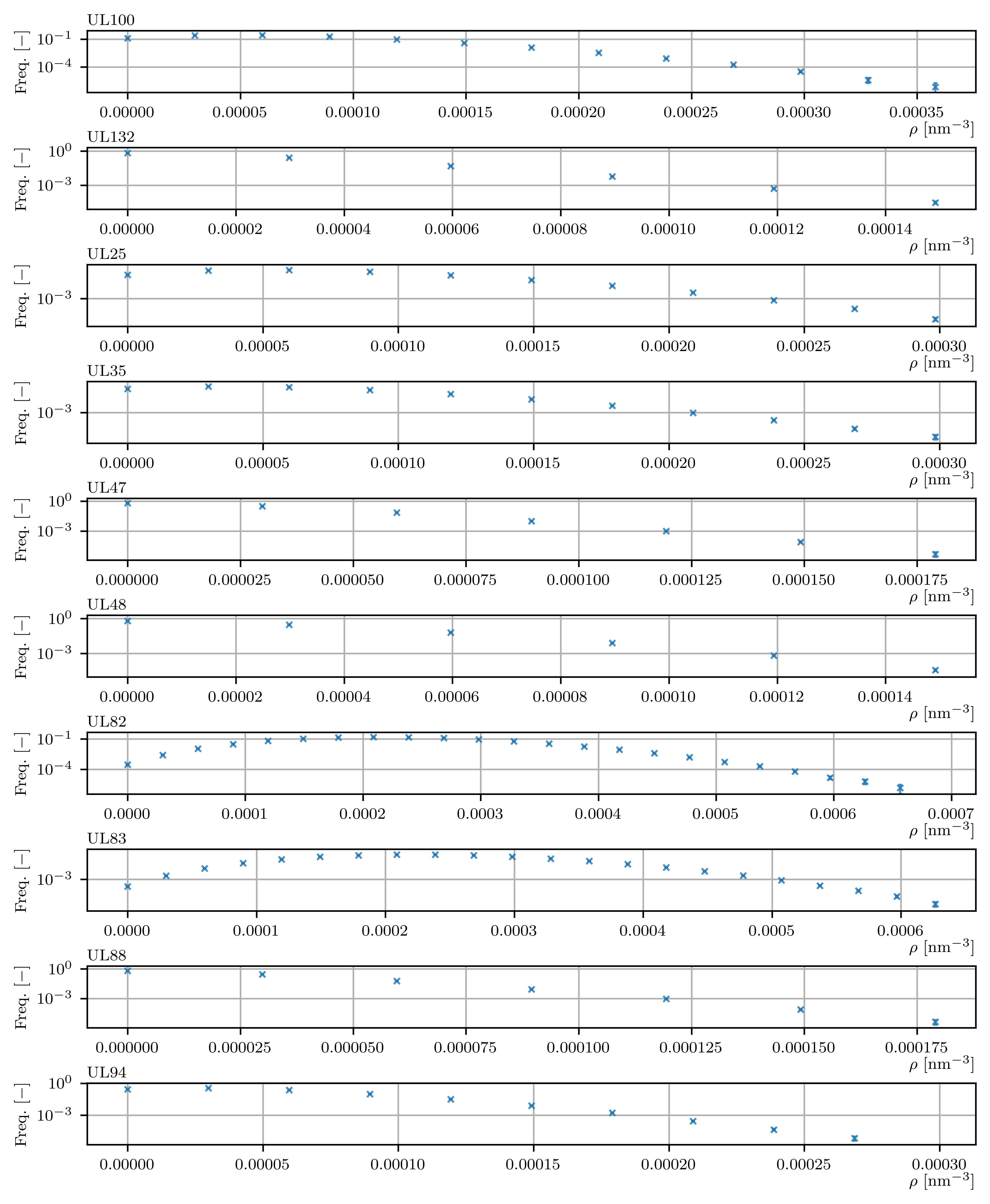}
    \caption{\label{fig:FlucRhoSinglePhase}Distributions of the local density in the homogeneous reference system. The species type is specified in the top-left corner of each plot. Due to the logarithmic axes, confidence intervals including zero or negative values could not be fully displayed. For most other points confidence intervals are comparable to the markersize.}
\end{figure*}

\begin{figure*}
    \centering
    \includegraphics[height=7.8in]{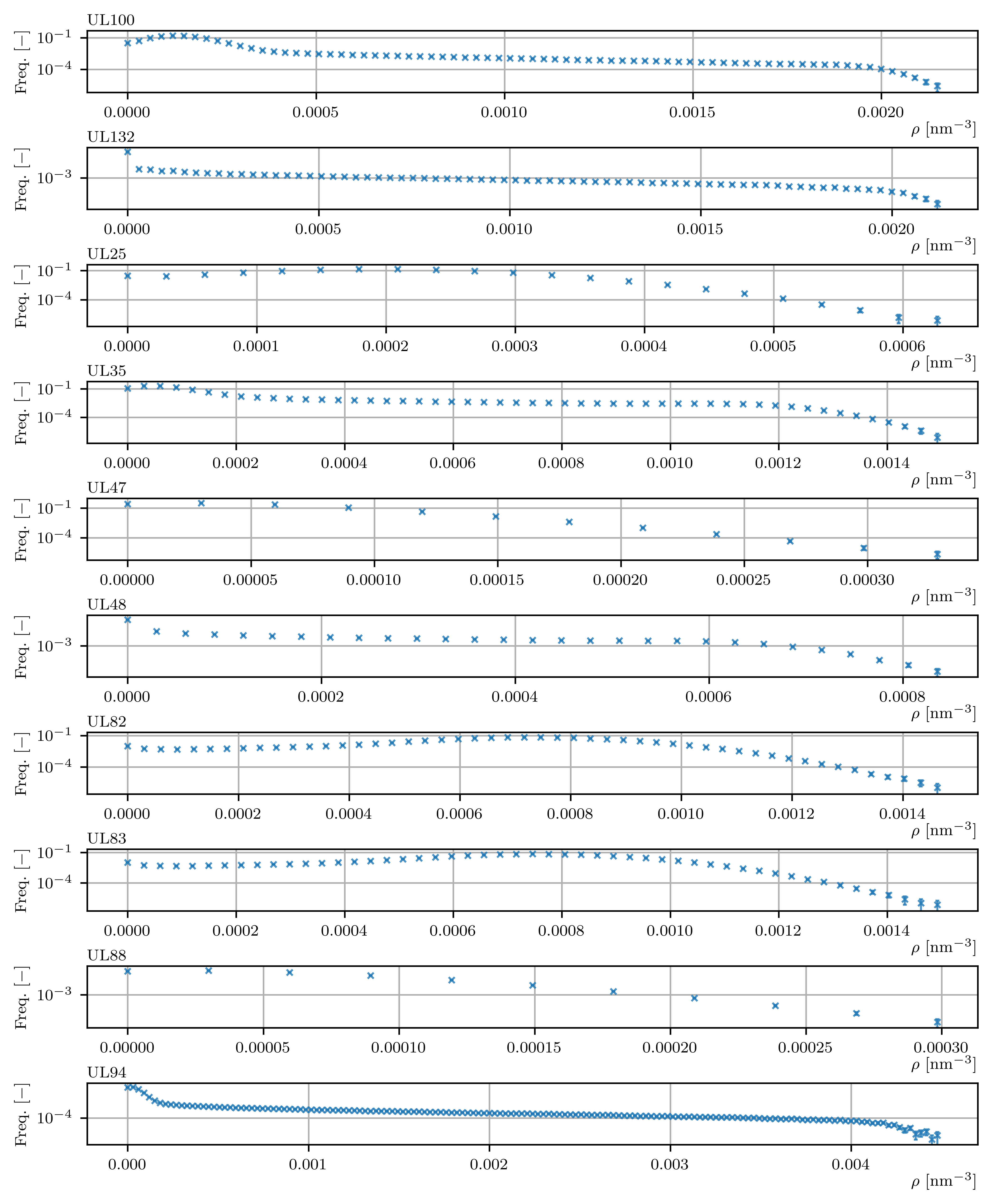}
    \caption{\label{fig:FlucRhoMultiPhase}Distributions of the local density in the multiphase reference system. The species type is specified in the top-left corner of each plot. Due to the logarithmic axes, confidence intervals including zero or negative values could not be fully displayed. For most other points confidence intervals are comparable to the markersize.}
\end{figure*}

To further characterize the thermodynamic state of both systems, fluctuations of the local particle density were analyzed. This was done by calculating the local density in $100$~randomly placed spheres (radius~$r=20~\text{nm}$) for each frame of the trajectory. The resulting histograms are shown in Figure~\ref{fig:FlucRhoSinglePhase} and~\ref{fig:FlucRhoMultiPhase} for one representative simulation of each system.

In the homogeneous reference system (Figure~\ref{fig:FlucRhoSinglePhase}), the density distributions for all particle types are comparatively narrow and roughly proportional to their respective particle numbers~$N_a$, consistent with uniform mixing. Maximum local densities remain modest, with the most abundant species (UL82, UL83) reaching approximately~$0.0006~\text{nm}^{-3}$, while less abundant species show correspondingly lower peak densities.

In striking contrast, the multiphase reference system for which results are displayed in Figure~\ref{fig:FlucRhoMultiPhase} exhibits dramatically broader density distributions with significantly higher maximum values. Most notably, UL94 reaches local densities of up~to~$0.004~\text{nm}^{-3}$ compared to~$0.00025~\text{nm}^{-3}$ in the homogeneous system. Similarly, UL100 and UL132 exhibit similar maximum densities of approximately~$0.002~\text{nm}^{-3}$, compared to~$0.0004~\text{nm}^{-3}$ in the homogeneous case. These elevated densities directly reflect the formation of concentrated phases enriched in these particle types.

It should be noted that multimodal distributions would, in principle, be expected for a multiphase system. However, with a probe radius of~$r=20~\text{nm}$ and box length of approximately~$140~\text{nm}$, many sampled volumes inevitably span phase boundaries. This is particularly true given the relatively high surface-to-volume ratio of the condensed phases in this comparatively small system. Consequently, the observed distributions are smeared into broad unimodal forms that nevertheless exhibit dramatically different ranges compared to the homogeneous system.

Together, these results demonstrate that the homogeneous system exhibits spatially uniform densities with narrow fluctuations, while the multiphase system displays clear spatial segregation and substantial density fluctuations, confirming the formation of multiple coexisting phases with distinct compositions.

\clearpage
\paragraph*{Uncertainty Analysis}
To ensure the reliability of the obtained pair numbers, three independent simulations were conducted for each reference system using \textit{normal} initialization (Section~\ref{sec:Exp:SimProtocol}). Since multiphase systems can exhibit metastability, three additional multiphase simulations were performed. In these simulations, the starting positions were not minimized in energy and no external force was applied during equilibration. This approach, which will be referred to as \textit{non-minimized} initialization was meant to ensure that the formation of condensed phases was thermodynamically driven, and not biased by the initialization protocol.

The algorithms in Sections~\ref{sec:Results:Homogeneous} and~\ref{sec:Results:InHomogeneous} were tested using the pair numbers obtained as the average over the three normally initialized simulations, which will be referred to as the \textit{ideal} pair numbers. To verify that no systematic deviations from these pair numbers occurred in any reference simulation, the~$\mathcal{L}$ (eqn.~\ref{eqn:MSLogLoss}) between the ideal pair numbers~$z_i^\ast$ and the pair numbers obtained in each simulation~$z_i$ was estimated for each run. The results are here listed in Table~\ref{tab:rMADSingleResults} and~\ref{tab:rMADMultiResults}.

\begin{table}
    \caption{\label{tab:rMADSingleResults}Mean Squared Logarithmic Deviation~$\mathcal{L}$ (eqn.~\ref{eqn:MSLogLoss}) of the homogeneous reference simulations. The metric was recorded relative to the average across all three simulations.}
    \centering
    \begin{tabular}{c|c}
         Run & $\mathcal{L}$ [-] \\ \hline
         \vphantom{$\left(0.000040^{+53}_{-25}\right)^a$} 1 & $0.000040^{+53}_{-25}$ \\[2pt]
          2 & $0.000032^{+39}_{-18}$ \\[2pt]
          3 & $0.000044^{+59}_{-26}$
    \end{tabular}
\end{table}

\begin{table}
    \caption{\label{tab:rMADMultiResults}Mean Squared Logarithmic Deviation~$\mathcal{L}$ (eqn.~\ref{eqn:MSLogLoss}) of the multiphase reference simulations. The metric was recorded relative to the average pair numbers from the three normally initialized simulations. The uncertainty quantification procedure outlined in Section~\ref{sec:Exp:UQprotocol} did not converge for the third non-minimized simulation such that the CIs reported here might be underestimated.}
    \centering
    \begin{tabular}{lc|c}
        Initialization & Run & $\mathcal{L}$ [-] \\ \hline
         \vphantom{$\left(0.0028^{+75}_{-20}\right)^a$} Normal & 1 & $0.0028^{+75}_{-20}$ \\[2pt]
          & 2 & $0.014^{+31}_{-14}$ \\[2pt]
          & 3 & $0.006^{+15}_{-06}$ \\[2pt] \hline
         \vphantom{$\left(0.025^{+25}_{-21}\right)^a$} Non-Minimized & 1 & $0.025^{+25}_{-21}$ \\[2pt]
          & 2 & $0.007^{+16}_{-07}$ \\[2pt]
          & 3 & $0.009^{+27}_{-09}$
\end{tabular}
\end{table}

The dynamics of the homogeneous system were sampled very efficiently, yielding highly accurate pair number estimates. Accordingly, deviations from the ideal pair numbers remained relatively small, and very low $\mathcal{L}$~estimates were found. As listed in Table~\ref{tab:rMADSingleResults}, the strongest deviations were found to result in a value of~$\mathcal{L}=0.000044^{+59}_{-26}$, indicating rather precise reference data. Overall, the obtained deviations indicate a "ground-truth" value of~$\mathcal{L}\approx4\times10^{-5}$.

In contrast, the multiphase system exhibited significantly poorer sampling, resulting in considerably larger deviations. As noted in Table~\ref{tab:rMADMultiResults}, the normally initialized simulations yielded $\mathcal{L}$~estimates as high as~$\mathcal{L}=0.014^{+31}_{-14}$. Although this presents a 200-fold increase compared to the homogeneous system, it corresponds to a mean relative absolute deviation of approximately~$10~\%$, which is acceptable. Additionally, it should be noted that the observed deviations in~$\mathcal{L}$ might seem substantial, however, all obtained CIs are in agreement around a value of~$\mathcal{L}\approx0.01$, and~$\mathcal{L}$ is the mean \textit{squared} logarithmic deviation, making it rather sensitive to fluctuations.

Table~\ref{tab:rMADMultiResults} also shows the results obtained for non-minimized simulations. Here, it can be seen that all $\mathcal{L}$~estimates are in agreement, irrespective of initialization protocol. While the non-minimized simulations yielded slightly higher $\mathcal{L}$~estimates, reaching up~to~$\mathcal{L}=0.025^{+25}_{-21}$ for simulation one, all CIs overlap and the observed deviations are rather small in absolute terms. This suggests that the observed deviations are statistical rather than systematic, indicating that the choice of initialization does not bias the final equilibrium pair numbers. Similarly, while the uncertainty quantification protocol outlined in Section~\ref{sec:Exp:UQprotocol} did not fully converge for the third non-minimized simulation, suggesting that the CI reported for this simulation might be an underestimate, the observed value is in good agreement with all others, such that this is likely not a large issue either. Lastly, given the obtained results, it appears that the chosen simulation parameters yield deviations leading to a value of~$\mathcal{L}\approx0.01$ which can be considered the "ground-truth"~$\mathcal{L}$.

\bibliography{mybib}

\end{document}